\documentclass[journal,10pt,twocolumn,]{IEEEtran}

\usepackage{cite}
\usepackage{amsmath, amssymb}

\DeclareMathOperator*{\argmin}{arg\,min}
\usepackage{graphicx}
\usepackage{enumitem}
\usepackage{bm}
\usepackage{mathtools}
\DeclarePairedDelimiter\abs{\lvert}{\rvert}

\usepackage{amsfonts}
\usepackage{stackengine}
\usepackage{xcolor}
\usepackage{algorithm}
\usepackage[export]{adjustbox}
\usepackage[noend]{algpseudocode}
\stackMath
\usepackage{multirow} % Required for multirows

\usepackage{relsize}
\usepackage{bm}
\usepackage{soul}
\usepackage{subcaption}
\let\latexchi\chi
\makeatletter
\renewcommand\chi{\@ifnextchar_\sub@chi\latexchi}
\newcommand{\sub@chi}[2]{% #1 is _, #2 is the subscript
	\@ifnextchar^{\subsup@chi{#2}}{\latexchi^{}_{#2}}%
}
\newcommand{\subsup@chi}[3]{% #1 is the subscript, #2 is ^, #3 is the superscript
	\latexchi_{#1}^{#3}%
}
\makeatother
\def\mathclap#1{\text{\hbox to 0pt{\hss$\mathsurround=0pt#1$\hss}}}
\newcommand{\raisedchi}{\raisebox{\depth}{\(\chi\)}}

\newcommand{\chimatrix}{\begin{bmatrix}
		\overline{\raisedchi}_R \\
		\overline{\raisedchi}_I
\end{bmatrix}}

\begin{document}
	
	\title{Unrolled Optimization with Deep Learning-based Priors for Phaseless Inverse Scattering Problems}
	
	\author{Samruddhi~Deshmukh,
		Amartansh~Dubey,~\IEEEmembership{Graduate Student Members,~IEEE},
		Ross~Murch,~\IEEEmembership{Fellow,~IEEE}
		% <-this % stops a space
		\thanks{This work was supported by the Hong Kong Research Grants Council with the Collaborative Research Fund C6012-20G.}
		\thanks{Samruddhi Deshmukh and Amartansh Dubey are with the Department of Electronic and Computer Engineering, Hong Kong University of Science and Technology (HKUST), Hong Kong (e-mail: ssdeshmukh@connect.ust.hk; adubey@connect.ust.hk).}
		\thanks{Ross Murch is with the Department of Electronic and Computer Engineering and the Institute of Advanced Study, Hong Kong University of Science and Technology (HKUST), Hong Kong (e-mail: eermurch@ust.hk).}
		% <-this % stops a space
	}

	\maketitle

	\begin{abstract}
		Inverse scattering problems, such as those in electromagnetic imaging using phaseless data (PD-ISPs), involve imaging objects using phaseless measurements of wave scattering. Such inverse problems can be highly non-linear and ill-posed under extremely strong scattering conditions such as when the objects have very high permittivity or are large in size. In this work, we propose an end-to-end reconstruction framework using unrolled optimization with deep priors to solve PD-ISPs under very strong scattering conditions. We incorporate an approximate linear physics-based model into our optimization framework along with a deep learning-based prior and solve the resulting problem using an iterative algorithm which is unfolded into a deep network. This network not only learns data-driven regularization, but also overcomes the shortcomings of approximate linear models and learns non-linear features. More important, unlike existing PD-ISP methods, the proposed framework learns optimum values of all tunable parameters (including multiple regularization parameters) as a part of the framework. Results from simulations and experiments are shown for the use case of indoor imaging using 2.4 GHz phaseless Wi-Fi measurements, where the objects exhibit extremely strong scattering and low-absorption. Results show that the proposed framework outperforms existing model-driven and data-driven techniques by a significant margin and provides up to 20 times higher validity range. 
	\end{abstract}
	
	\IEEEpeerreviewmaketitle
	
	\section{Introduction}
	\label{Sec_Introduction}
	
	Inverse scattering problems (ISPs) in electromagnetic imaging can be highly non-linear and ill-posed under strong scattering conditions and have therefore found limited use in fields such as large scale microwave imaging and indoor radio imaging. The non-linearity and ill-posedness of ISPs can vary based on factors such as the size and relative permittivity ($\epsilon_r$) of objects, the incident wavelength ($\lambda_0$), and the number and type of measurements available. They also increase significantly when the measurements are phaseless \cite{chen2018computational, xu2020deep, chen2010subspace}. Thus, PD-ISPs (Phaseless Data ISPs) are more challenging to solve than FD-ISPs (Full Data ISPs).
	
	\subsection{ISP with Phaseless Data (PD-ISP)}
	\label{Sec_PD_ISP}
	Practical solutions of PD-ISPs would be extremely useful, especially for applications such as indoor imaging using microwave frequencies where it is difficult to collect accurate phase measurements. Removing the need for phase can significantly reduce the cost and complexity of the measurement system and remove the need for synchronization between measurement nodes. 
	
	There has been limited research in solving PD-ISPs, mainly because they are highly non-linear and ill-posed and there is no exact formulation for its forward model \cite{chen2010subspace, xu2020deep}. Several linear and non-linear techniques have been proposed, and since PD-ISPs are inherently non-linear, the non-linear techniques achieve state-of-the-art performance in terms of validity range. However, these techniques are computationally expensive and are sensitive to experimental noise and errors, and therefore need controlled environments such as anechoic chambers to minimize noise and multi-path reflections from clutter (such as nearby objects, walls, ceiling and floor). On the other hand, linear models such as the phaseless Rytov Approximation have a lower validity range, but they provide more stable reconstructions even with imperfect experiment data. Thus, there is a trade-off between linear and non-linear models when it comes to handling non-ideal experiment data and achieving a better validity range.
	
	In our recent work \cite{dubey2022xPRA}, we propose a linear model denoted as the extended Phaseless Rytov Approximation in Lossy Media (xPRA-LM) which extends this trade-off. In terms of shape reconstruction, xPRA-LM can handle objects with extremely large permittivity and sizes. However, in terms of estimating permittivity, it fails under very strong scattering conditions as it relies on a linear approximation. 
	
	Motivated by the practical feasibility and the higher validity range of linear models such as xPRA-LM, this work aims to develop a deep learning framework that incorporates the physics-based xPRA-LM model to achieve a higher validity range. This can be achieved by learning the non-linear relationship between measurements and the object permittivity profile, while maintaining the ability to handle imperfect experiment data. 
	
	In the next subsection we provide a brief overview of existing deep learning techniques to solve ill-posed inverse problems. 

	\subsection{Deep Learning for Inverse Problems}
	\label{Sec_DLIP}
	Consider a general inverse problem of the form
	\begin{equation}
		\label{eq_generalISP}
		\min_{x} f(\bm {y}-{\bm A}(\bm {x})),
	\end{equation}
	where $\bm{y}$ is an $M \times 1$ measurement vector, $A$ is the $M \times N$ forward model that can operate linearly or non-linearly on the $N\times 1$ variable $x$ (which we wish to retrieve) and $f$ is the penalty function. The straightforward looking problem, (\ref{eq_generalISP}) can be difficult to solve due to three reasons: 1) the ill-posedness arising from scarcity of measurements ($M\ll N$) and noisy measurements, 2) the non-linear relation between $A$ and $x$, and, 3) the unavailability or inadequate knowledge of the underlying forward model $\bm{A}$. Inverse problems of the form (\ref{eq_generalISP}) in different fields can have one or more of these problems. For example, inverse problems in medical imaging are ill-posed, but the underlying model is usually known and is linear. In classical inverse scattering, FD-ISPs are ill-posed and non-linear, but the underlying non-linear model is known. However, PD-ISPs are more difficult to solve since they have all three problems - they are ill-posed, non-linear and the exact underlying model is unknown.
	
	Existing deep learning (DL) techniques solve inverse problems by tackling one or more of the aforementioned issues. The approaches can be  broadly categorized as follows:
	\subsubsection{\textbf{Direct Inversion (DI)}}
	DI techniques are fully data-driven/black-box approaches that attempt to directly map measurements $y$ to reconstructions $x$ using a DL network. The measurements, which are typically 1D often need to be mapped to 2D or even 3D reconstructions. Also, since these techniques do not explicitly use any model information, the network needs to learn the forward model $A$ entirely from training data. Since these techniques completely rely on data to learn to tackle the ill-posedness and non-linearity and to estimate the forward model, they are highly data-intensive and lead to poor generalization.
	
	\subsubsection{\textbf{Model-based Deep Learning}}
	These techniques use a fully/partially known forward model $A$ to solve the inverse problem which is then enhanced using DL. They can be further classified into two categories:
	\begin{enumerate}
		
		\item \textbf{Model-assisted Deep Learning (MADL)}: In MADL techniques, a regularized pseudo-inverse of a linearized forward model is used to provide an approximate reconstruction, which is then processed using DL networks to obtain an enhanced final reconstruction \cite{jin2017deep, xu2020deep, deshmukh2022physics, sanghvi2019embedding}. Therefore, the DL network in MADL learns to tackle ill-posedness and non-linearity, while the forward model is known. These techniques are less data-intensive as compared to DI and provide better generalization due to the use of the forward model. 
		
		\item \textbf{Model-guided Deep Learning (MGDL)}: These techniques rely on the forward model $A$ more strongly by incorporating it into an optimization framework or network architecture. A recent MGDL approach uses data-driven DL priors on the reconstructions while minimizing $f(y-A(x))$ as opposed to using conventional handcrafted priors such as  Tikhonov, LASSO, or TV. These data-driven priors can be further categorized based on their training mechanism as:
		\begin{enumerate}
			\item \textbf{Plug and Play priors}  \cite{zhang2017learning, rick2017one, meinhardt2017learning} where a DL network is first trained to learn structure in $x$, following which it is plugged into the objective in (\ref{eq_generalISP}). The resulting objective is solved using iterative optimization techniques.
			
			\item \textbf{Priors trained during optimization} \cite{diamond2017unrolled, adler2017solving, schlemper2017deep} where the deep prior is a part of the objective function from the beginning and it is trained while the optimization problem is solved. This is usually performed using unrolled optimization techniques. Learning deep priors through unrolled optimization leads to better generalization as compared to using plug and play priors and also removes the need to tune parameters of the optimization algorithm. 		
		\end{enumerate}
		
	\end{enumerate}

Based on the characteristics of the DL approaches described above we utilize the deep prior approach to provide enhanced solutions to our PD-ISP.
	
	\subsection{Motivation and Contributions}
	\label{Sec_Contributions}
	Most of the research on deep priors has been performed in the fields of medical imaging and image denoising or enhancement. Since the underlying forward model in these problems is linear in most cases, the key focus of most MGDL techniques is to reduce error caused by modeling errors, noise and ill-posedness. PD-ISPs on the other hand are different and in this work we attempt to solve PD-ISP of the form
	\begin{equation}
		\label{eq_generalPD_ISP}
		\min_{x} f(|\bm {y}|-\tilde{\bm A}_p \bm {x}),
	\end{equation}
	where we wish to reconstruct the $N\times1$ variable $x$ with an $M\times 1$ magnitude only measurement vector $|{\bm y}|$. The key difference is that the $M\times N$ matrix $\tilde{\bm A}_p$ here is a linear forward model that approximates an unknown underlying forward model $\bm{A}$ which is highly non-linear. As mentioned in Section \ref{Sec_PD_ISP}, the extent of the non-linearity of the actual forward model also depends on values of the elements of $x$ which are a function of the relative permittivity $\epsilon_r$ of the object. As $\epsilon_r$ increases, the non-linearity of the PD-ISP increases and the accuracy with which $\tilde{\bm A}_p$ approximates $\bm{A}$ decreases.
	
	To the best of our knowledge, there have been two attempts to solve PD-ISPs using deep learning \cite{xu2020deep, deshmukh2022physics}, both of which use the MADL approach to process noisy reconstructions obtained from a regularized pseudo-inverse of the forward model. However, there are a few issues that these techniques do not address. \cite{xu2020deep} uses a non-linear iterative forward model to generate the initial reconstructions, which is computationally expensive and limits experimental utility since non-linear models can be sensitive to experimental errors. Also, results are only demonstrated for moderate scattering conditions $(\epsilon_r\le3)$ whereas typical objects around us have $2<\epsilon_r<77$ (at frequencies around 2.4 GHz), making it a useful contribution but not yet widely useful. \cite{deshmukh2022physics} on the other hand uses a linear model to generate initial reconstructions and demonstrates results for strong scattering, but is not formulated for the lossy medium. Furthermore, both \cite{xu2020deep} and \cite{deshmukh2022physics} need a large amount of data to train their deep learning networks in order to efficiently denoise the initial reconstructions. Also, both these techniques need a trial-and-error based tuning of a critical hyperparameter to regularize the inversion of the ill-conditioned forward model and generate the initial reconstruction. This is not practical since this parameter also needs to be tuned during inference and the results depend heavily on it. MADL approaches are also not highly interpretable since they typically use black-box DL networks to process the initial reconstruction. 
	
	In this work we present an end-to-end unrolled optimization framework with deep priors which is equipped to tackle the severe non-linearity and ill-posedness of PD-ISPs and handle the aforementioned problems. The key contributions can be summarized as follows: 
	\begin{enumerate}[leftmargin=0.45cm]
		\item \textbf{Forward Model:} We use our recently proposed extended phaseless Rytov approximation \cite{dubey2022xPRA} in lossy media (xPRA-LM) as a phaseless linear forward model $\tilde{\bm A}_p$ in our framework. xPRA-LM is shown to have the highest validity range in handling strong scattering among all PD-ISP models and can also handle experimental errors due to its linear formulation. It is also the only PD-ISP model which can be formulated to include background subtraction to remove multipath reflections (from clutter, walls, celing, floor) from experimental data. We incorporate xPRA-LM into our proposed end-to-end reconstruction framework.
		
		\item \textbf{End-to-end unrolled optimization framework with deep priors} which:
		\begin{enumerate}
			\item Uses data-driven regularization in the form of deep learning based priors for the xPRA-LM inverse problem while also learning the underlying non-linearity which xPRA-LM cannot model,
			
			\item Uses conventional multi-parameter regularization in the objective in addition to the data-driven regularization to pre-condition xPRA-LM, thus tackling the ill-posedness of the xPRA-LM inverse problem, and
			
			\item Learns all optimum parameter values (including  multiple regularization parameters and learning rate of the optimization algorithm) during training, removing the need to tune parameters manually.
			
		\end{enumerate}
		
	\end{enumerate}

	We demonstrate the performance of the proposed framework using simulation examples and experimental examples where measurement data is collected in a realistic indoor environment using Wi-Fi devices.	
	
	\subsection{Notation}
	\label{Sec_Organization}
	We use $\overline{\overline{X}}$ and $\overline{X}$ to denote the matrix and vector form of discretized parameter $X$ respectively. Lower case boldfaced letters represent position vectors and italic letters represent scalar parameters.
	
	\section{Problem Formulation}
	\label{Sec_ProblemSetup}
	
\begin{figure}[h]
		\centering
		\includegraphics[width=2.2in]{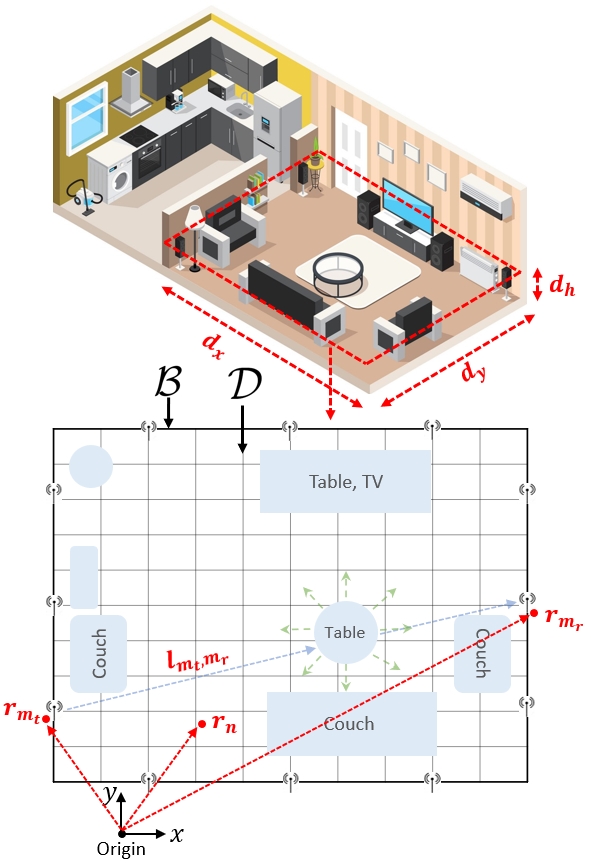}
		\caption{The DoI with wireless transceiver nodes at its boundary $ {\mathcal{B}}$ and dimensions $d_x \times d_y$ at cross-sectional height $d_h$. The transmitter $m_t$ and receiver $m_r$ are located at ${\bf r}_{m_t} \in {\mathcal{B}}$ and ${\bf r}_{m_r} \in {\mathcal{B}}$ respectively. The location of the $n^{th}$ grid is denoted as ${\bf r}_n  \in {\mathcal{D}}$.}
		\label{RTInetwork2}
	\end{figure}

	Consider a Domain of Interest (DoI) ${\mathcal{D}}\subset \mathbb{R}^{2}$ situated in an indoor region of a building as shown in Fig. \ref{RTInetwork2}. It has dimensions $d_x \times d_y$ m$^2$ and has 2.4 GHz Wi-Fi transceiver nodes placed at its boundary (denoted as $\mathcal{B} \subset \mathbb{R}^{2}$). There are in total $M$ transceiver nodes, each of which transmit and receive signals to acquire Received Signal Strength (RSS) measurements of links between the nodes. These nodes cannot transmit and receive at the same time. Therefore, the total number of wireless links is equal to $L=M(M-1)/2$ excluding the reciprocal links and self-measurement. The $m^{th}$ node is located at ${\bf r}_m \in \mathcal{B}$ and $l_{m_t,m_r}$ denotes the link between the transmitting node $m_t$ (at ${\bf r}_{m_t}$) and the receiving node $m_r$ (at ${\bf r}_{m_r}$). For the remainder of this work, we use the subscripts $m_t$ and $m_r$ to refer to the transmitter and receiver respectively for all relevant quantities. 
	
	We can approximate the environment in Fig. \ref{RTInetwork2} as a 2D electromagnetic problem where we consider the DoI to be a planar cross-section parallel to the floor at a height $d_h$. This 2D approximation is valid since we use directional antennas and background subtraction \cite{dubey2022xPRA} to reduce the effect of scattering from the floor, ceiling and other clutter outside the DoI. We discretize this 2D DoI into $N = n_x \times n_y$ rectangular grids, each of size $\Delta d_x \times \Delta d_y$, where ${\bf r}_n$ denotes the location of the $n^{th}$ grid (see Fig. \ref{RTInetwork2}).
	
	Let us consider the $l^{th}$ wireless link with the source at $\bm{r}_{m_t}$ and the receiver at $\bm{r}_{m_r}$ as shown in Fig. \ref{RTInetwork2}. The total field $E_{m_t}(\bm{r})$ at the receiver ($\bm{r} = \bm{r}_{m_r}$) or inside any grid in the DoI ($\bm{r} = \bm{r}_{n}$) is
	\begin{equation}
		\label{Eq_Hzfree}
		\begin{aligned}
			E_{m_t}(\bm{r}) = E^i_{m_t}(\bm{r}) + E^s_{m_t}(\bm{r}),
		\end{aligned}
	\end{equation}
	where $E^i_{m_t}(\bm{r})$ is the free-space incident field and $E^s_{m_t}(\bm{r})$ is the scattered field in the presence of scattering objects. The total field at the receiver $E_{m_t}(\bm{r}_{m_r})$ can be expressed in an exact form in terms of the total field inside the DoI $E_{m_t}(\bm{r}_n)$ and the permittivity profile of the DoI $\epsilon_r({\bf r}_n)$ using a volume source integral (VSI) formulation as
	\begin{equation}
		\label{Eq_VSI}
		\begin{aligned}
			E_{m_t}(\bm{r}_{m_r}) = E^i_{m_t}(\bm{r}_{m_r}) + k_0^2 \int_{\mathcal{D}} g(\bm{r}_{m_r}, \bm{r}_{n}) \chi_{\epsilon}(\bm{r}_n) E(\bm{r}_{n}) d\bm{r}_{n},
		\end{aligned}
	\end{equation}
	where $\chi_{\epsilon}(\bm{r}_n) = \epsilon_r(\bm{r}_{n})-1$ is the permittivity contrast and $g$ is the homogeneous Green's function. The VSI equation in (\ref{Eq_VSI}) is also known as the Lipmann-Schwinger equation and provides an exact description of wave scattering \cite{Mittra1998, chen2018computational}.
	
	Solving VSI as an inverse problem implies solving it to estimate the contrast profile $\chi(\bm{r}_n)$ of the DoI given measurements of $E$ (on $\mathcal{B}$). This is known to be a non-linear and ill-posed problem since there are two unknowns inside the integral ($\chi(\bm{r}_n)$ and  $E(\bm{r}_n)$), both of which need to be solved for a 2D DoI with only 1D measurements of the total field (along the measurement boundary $\mathcal{B}$). Also, VSI in (\ref{Eq_VSI}) needs both the phase and magnitude and is therefore an FD-ISP. There is no exact reformulation of VSI as a PD-ISP. Therefore, in the next section, we provide a brief description of an approximate linear PD-ISP model derived from FD-ISP in (\ref{Eq_VSI}).
	
	\section{Proposed Phaseless Approximate Model For Model Guided Deep Learning}
	\label{Sec_xPRA_LM}
	In this section we approximate VSI, (\ref{Eq_VSI}), using our recently proposed xPRA model \cite{dubey2022xPRA} derived by correcting the well known Rytov approximation (RA) and reformulate it as a linear approximate PD-ISP where the contrast is linearly related to RSS values. 
	
	In RA, instead of defining the total field as $E = E_i+E_s$, it is defined as
	\begin{equation}
		\label{Eq_RA}
		\begin{aligned}
			E_{m_t}(\bm{r}) = E^i_{m_t}(\bm{r})  e^{\phi^s_{m_t}(\bm{r})},
		\end{aligned}
	\end{equation}
	where the complex phase ${\phi^s_{m_t}}$ represents the phase and log amplitude deviations of the total field from the incident field as the field wavefront passes through the scattering media. However, RA is only valid for extremely weak scattering ($|\epsilon_r|\approx 1$) with limited practical applications. 
	Our recent work \cite{dubey2022xPRA} provides corrections to RA using a high frequency approximation theory in lossy media to also handle strong scattering conditions. Combining this new corrected contrast with RA leads to the extended Rytov Approximation for lossy media (xRA-LM) \cite{dubey2022xPRA} which can be written as 
	\begin{equation}
		\label{Eq_RIfinal2}
		\begin{aligned}
			E_{m_t} & (\bm{r}_{m_r})  = E^i_{m_t}(\bm{r}_{m_r}) \ \cdot \\  & \exp\left( \frac{k_0^2}{E^i_{m_t}(\bm{r}_n)} \int_{\mathcal{D}}  g(\bm{r}_{m_r}, \bm{r}_n)  \raisedchi_{m_t}(\bm{r}_n)  E^i_{m_t}(\bm{r}_n) d\bm{r'}^2 \right),
		\end{aligned}
	\end{equation} 
	where $\raisedchi_{m_t}$ is the contrast given as
	\begin{equation}
		\label{Eq_RIfinal2a}
		\begin{aligned}
			\raisedchi_{m_t}(\bm{r}_n) & =2 (\sqrt{\epsilon_R(\bm{r}_n)} \cos\theta^{s}_{m_t}(\bm{r}_n) -1)	\\ & + j  \frac{\epsilon_I(\bm{r}_n)}{\sqrt{\epsilon_R(\bm{r}_n) - \sin^2\theta^i_{m_t}(\bm{r}_n)}} \cos\theta^i_{m_t}(\bm{r}_n),
		\end{aligned}
	\end{equation} 
	where $\epsilon_R(\bm{r}_n)$ and $\epsilon_I(\bm{r}_n)$ are the real and imaginary parts of relative permittivity $\epsilon_r(\bm{r}_n) = \epsilon_R(\bm{r}_n)+j \epsilon_I(\bm{r}_n)$. $\theta^s_{m_t}(\bm{r}_n)$ and $\theta^i_{m_t}(\bm{r}_n)$ are the scattering angle and the incident angle respectively \footnote{The incident angle is the angle made by incident wave at the grid $\bm{r}_n$ of the DoI and the scattering angle represents how much this incident wave gets scattered (refracted/reflected) from its original path.}. Unlike VSI, (\ref{Eq_RIfinal2}) is now a linear inverse problem where the unknown contrast $\chi_{m_t}(\bm{r}_n)$ is linearly related to measurements $E_{m_t}(\bm{r}_{m_r})$. Note that $\raisedchi_{m_t(\bm{r}_n)}$ is also a function of the location of the transmitting node due to its dependence on $\theta^s_{m_t}(\bm{r}_n)$ and $\theta^i_{m_t}(\bm{r}_n)$, where $\theta^s_{m_t}(\bm{r}_n)$ itself depends on $\theta^i_{m_t}(\bm{r}_n)$.

	\subsection{Phaseless Form}
	\label{Sec_phaseless}
	One of the biggest advantages of the exponential forms of FD-ISP in (\ref{Eq_RIfinal2}) is that it can be transformed into a phaseless form (PD-ISP). This can be done by multiplying (\ref{Eq_RIfinal2}) by its conjugate and taking logarithm on both sides to obtain
	\begin{equation}
		\label{Eq_RytovInt}
		\begin{aligned}
			& \Delta P_{m_t}  (\bm{r}_{m_r}) =  \\ &  C_0 \cdot \operatorname{Re}\bigg(\frac{k^2}{E^i_{m_t}(\bm{r}_{m_r})}  \int_{\mathcal{D}} g(\bm{r}_{m_r}, \bm{r}_n)   \raisedchi_{m_t}(\bm{r}_n) E^i_{m_t}(\bm{r}_n) d\bm{r'}^2\bigg),
		\end{aligned}
	\end{equation}
	where the constant $C_0 = 20\log_{10}e$ and $\Delta P  (\bm{r}_{m_r})$ is the change in RSS values due to presence of the scatterer and can be written as
	\begin{equation}
		\begin{aligned}
			\Delta P_{m_t}(\bm{r}_{m_r}) [\text{dB}] &=  P_{m_t}(\bm{r}_{m_r}) [\text{dB}]- P_{m_t}^i(\bm{r}_{m_r})[\text{dB}] \\
			&=  20 \log_{10} \frac{|E_{m_t} (\bm{r}_{m_r})|}{|E_{m_t}^i (\bm{r}_{m_r})|},
		\end{aligned}
	\end{equation}
	where $P_{m_t}$ and $P_{m_t}^i$ are the total received power and the free-space incident power respectively. This means that to estimate the contrast profile $\raisedchi_{m_t}(\bm{r}_n)$, we only need phaseless measurements of the total field. Therefore, unlike (\ref{Eq_VSI}) and (\ref{Eq_RIfinal2}) which are FD-ISPs, (\ref{Eq_RytovInt}) is a PD-ISP. Also, (\ref{Eq_RytovInt}) is a linear inverse problem where the unknown $\raisedchi$ is linearly related to the change in RSS values ($P-P_i$).  We refer to this as the xPRA model in remainder of the paper. To the best of our knowledge, there are no other non-iterative linear PD-ISP models.
	
	%\subsection{Interpretations}
	Unlike exact models such as VSI where the contrast $\raisedchi_{\epsilon}$ is a linear function of permittivity $\epsilon_r$, the new contrast $\raisedchi_{m_t}$ in (\ref{Eq_RIfinal2a}) is now a non-linear function of $\epsilon_r$ and includes new distortion terms that depend on the scattering angle $\theta^s_{m_t}$ and the incident angles $\theta^i_{m_t}$.
	Therefore, even if we obtain the contrast profile by solving (\ref{Eq_RytovInt}), it will contain distortions that are not trivial to remove. Since $\theta^s_{m_t}(\bm{r}_n)$ itself depends on both $\theta^i_{m_t}(\bm{r}_n)$ and $\epsilon_r(\bm{r}_n)$, it causes significant distortion in $\operatorname*{Re}(\raisedchi_{m_t}(\bm{r}_n))$ and this distortion is different for different $\epsilon_r$ values. On the other hand, distortion in $\operatorname*{Im}(\raisedchi_{m_t}(\bm{r}_n))$ due to $\theta^i_{m_t}(\bm{r}_n)$ terms does not depend on $\epsilon_r(\bm{r}_n)$ since $\theta^i_{m_t}(\bm{r}_n)$ does not depend on $\epsilon_r(\bm{r}_n)$ and instead only depends on the transmitting node location. Hence, any distortions in $\operatorname*{Im}(\raisedchi_{m_t}(\bm{r}_n))$ will be the same no matter what the permittivity of the object is. The effect of this is shown in our recent work \cite{dubey2022xPRA} where the reconstruction of $\operatorname*{Re}(\raisedchi_{m_t}(\bm{r}_n))$ using xPRA is highly distorted due to $\theta^s_{m_t}(\bm{r}_n)$ term. On the other hand, reconstruction of $\operatorname*{Im}(\raisedchi_{m_t}(\bm{r}_n))$ using xPRA provides accurate shape reconstruction for very strong, low loss scatterers and helps distinguish between objects with different permittivity values. However, even reconstructing $\operatorname*{Im}(\raisedchi_{m_t}(\bm{r}_n))$ using xPRA has several limitations. Distortion occurs in the reconstructions because $\theta^i_{m_t}(\bm{r}_n)$ is not known or estimated. Also, xPRA cannot estimate $\operatorname*{Im}(\raisedchi_{m_t}(\bm{r}_n))$ and distinguish between objects when the relative permittivity becomes very large ($\epsilon_R(\bm{r}_n)>10$). Finally, it cannot accurately image objects with non-convex shapes.
	
	The limitations of xPRA are a result of using a linear approximation and the distortions due to the $\theta^i_{m_t}(\bm{r}_n)$ and $\theta^s_{m_t}(\bm{r}_n)$ terms. In this work, we aim to tackle these limitations of xPRA to enhance its validity. The idea is to combine xPRA with a deep learning framework that learns to reconstruct the spatial structure (such as non-convex shapes) and large contrast amplitude values of scatterers which cannot be reconstructed using xPRA alone, while also removing distortions in the reconstructions due to the $\theta^i_{m_t}(\bm{r}_n)$ and $\theta^s_{m_t}(\bm{r}_n)$ terms. 
	
	Our previous results \cite{dubey2022xPRA} showed that $\operatorname*{Im}(\raisedchi_{m_t}(\bm{r}_n))$ is closely related to the reconstruction of the ratio $\epsilon_I(\bm{r}_n)/\sqrt{\epsilon_R(\bm{r}_n)} = \delta(\bm{r}_n) \sqrt{\epsilon_R(\bm{r}_n)}$ where $\delta(\bm{r}_n)$ is the loss tangent.  While it cannot separately estimate $\epsilon_R$ and $\epsilon_I$ profiles of the DoI, the estimation of the $\delta \sqrt{\epsilon_R}$ profile provides an estimate of both the scattering ability ($\epsilon_R$) and the absorbing ability of the object under test. It can be thought of as a type of extinction coefficient. It is sufficient to obtain accurate shape reconstructions of extremely strong scatterers (even of non-convex shapes) and can also efficiently distinguish between objects of different permittivity values. 
	
	For the reasons described above, we propose to learn the $\delta(\bm{r}_n) \sqrt{\epsilon_R(\bm{r}_n)}$ ratio from $\operatorname*{Im}(\raisedchi_{m_t}(\bm{r}_n)))$  and  $\operatorname*{Re}(\raisedchi_{m_t}(\bm{r}_n)))$. We show later in the results section that our technique can provide extremely accurate reconstructions of $\delta(\bm{r}_n) \sqrt{\epsilon_R(\bm{r}_n)}$.

	\subsection{Imaging Model in Matrix-vector form}
	\label{Sec_xPRA-LM}
	Since the DoI is divided into $N = N_x \times N_y$ grids, we can replace the integral in the aforementioned equations with summation. Also, the xPRA model in (\ref{Eq_RytovInt}) is for a single wireless link. We can stack it for all $L = M(M-1)/2$ measurement links to obtain a linear system of equations
	\begin{equation}
		\label{Eq_discrete1}
		{\Delta \overline{ P} \approx \operatorname{Re}\big(\overline{\overline{G}} \  { \overline{\raisedchi}} \big)\ },
	\end{equation}
	where the contrast vector ${ \overline{\raisedchi}}\in \mathbb{C}^{N\times 1}$ contains the elements ${ {\raisedchi(\bm{r}_n)}}$ for all $N$ grids inside the DoI ($n = 1, 2,...,N$). Note that $\raisedchi_{m_t}(\bm{r}_n)$ in (\ref{Eq_RytovInt}) is a function of a single transmitting node $m_t$, whereas the vector $\overline{\raisedchi}$ in (\ref{Eq_discrete1}) is a single $N \times 1$ vector obtained using $L$ measurements due to illumination of DoI by all the transmitting nodes in the measurement setup. Therefore, $\overline{\raisedchi}$ can be considered to be a weighted linear combination of $\overline{\raisedchi}_{m_t}$ vectors for all measurement links and should only be a function of permittivity. It is from this term $\overline{\raisedchi}$ that we learn $\delta(\bm{r}_n) \sqrt{\epsilon_R(\bm{r}_n)}$. We do this by using data-driven regularization modeled as deep priors as shown later. Each element of the measurement vector $\Delta \overline{P} \in \mathbb{R}^{L\times 1}$ is $p_l= \Delta {P}({\bf r}^{ l}_{{m_r}}) = P({\bf r}^{ l}_{{m_r}})-P^i({\bf r}^{ l}_{{m_r}})$ [in dB]. The position vector ${\bf r}^{ l}_{{m_r}}$ gives the location of the receiver in the $l^{th}$ wireless link where the transmitter is at ${\bf r}^{ l}_{{m_t}}$. The xPRA kernel matrix ${\overline{{\overline{G}}}} \in \mathbb{C}^{L\times N}$ contains entries ${\overline{{\overline{G}}}}_{l,n} $ given as
	\begin{equation}
		\label{Eq_discrete2}
		{{\overline{{\overline{G}}}}_{l,n} =  \frac{C_0 k_0^2}{E^i({\bf r}^{ l}_{{m_r}})}\sum_{\forall n}  g({\bf r}^{ l}_{{m_r}}, \bm{r}_n)  E^i(\bm{r}_n) \Delta a},
	\end{equation}
	where $\Delta a$ is the area of each grid.

	We can further simplify (\ref{Eq_discrete1}) by manipulating the operator $\operatorname*{Re}$ to rewrite (\ref{Eq_discrete1}) as
	\begin{equation}
		\label{Eq_discrete3}
		{\Delta \overline{P} = \big[\ \operatorname{Re}({\overline{\overline{G}}}) \ \ \ -\operatorname{Im}({\overline{\overline{G}}}) \ \big]\ } \begin{bmatrix}
			{\operatorname{Re}({ \overline{\raisedchi}})} \\
			{\operatorname{Im}({ \overline{\raisedchi}}) }
		\end{bmatrix}.
	\end{equation}
	We can define our inverse problem as estimating $\operatorname{Re}({ \overline{\raisedchi}})$ and  $\operatorname{Im}({ \overline{\raisedchi}})$ given the measurements $\Delta \overline{P}$. Therefore, the model based objective function (which will be used in the next section) can be defined as
	\begin{equation}
		\label{Eq_discrete4}
		\min_{\chi_R, \chi_I} \abs*{\abs*{\Delta \overline{P} - \overline{\overline{\mathcal{G}}}\ 
				\begin{bmatrix}
					\overline{\raisedchi}_R \\
					\overline{\raisedchi}_I
		\end{bmatrix}}}^2_2,
	\end{equation}
	where 
	\begin{equation}
		%\label{Eq_discrete4}
		\begin{aligned}
			\overline{\overline{\mathcal{G}}} &= \left[ \operatorname{Re}({\overline{\overline{G}}}), \  \ -\operatorname{Im}({\overline{\overline{G}}}) \right] && \in \mathbb{R}^{L\times 2N}, \\
			\ \ \ \ \ \ \overline{\raisedchi}_R & =  \operatorname*{Re}({ \overline{\raisedchi}}), \ \ \overline{\raisedchi}_I =  \operatorname*{Im}({ \overline{\raisedchi}}) && \in \mathbb{R}^{N\times1}.
		\end{aligned}
	\end{equation}
	The matrix $\overline{\overline{\mathcal{G}}} \in \mathbb{R}^{L\times 2N}$ is an approximation to the underlying non-linear forward model. 
	
	As explained in the previous subsection, $\operatorname{Im}({ \overline{\raisedchi}})$ is more relevant than $\operatorname{Re}({ \overline{\raisedchi}})$ for reconstructing the $\delta \sqrt{\epsilon_R}$ profile since $\operatorname{Re}({ \overline{\raisedchi}})$ is distorted due to the presence of the unknown $\theta^s_{m_t}$ terms. In the next section we utilize our MGDL framework to reconstruct the $\delta \sqrt{\epsilon_R}$ profile of the DoI using both $\operatorname{Im}({ \overline{\raisedchi}})$ and $\operatorname{Re}({ \overline{\raisedchi}})$ so that all useful information is exploited. More specifically, since $\operatorname{Im}({ \overline{\raisedchi}})$ provides approximate reconstructions of the $\delta \sqrt{\epsilon_R}$ profile, we use $\operatorname{Im}({ \overline{\raisedchi}})$ as the primary input and $\operatorname{Re}({ \overline{\raisedchi}})$ as the secondary input in our MGDL framework. 
	
	\section{End-to-End Reconstruction Framework}
	\label{Sec_ProposedMethod}
	\begin{figure*}
		\centering
		\includegraphics[scale=0.45]{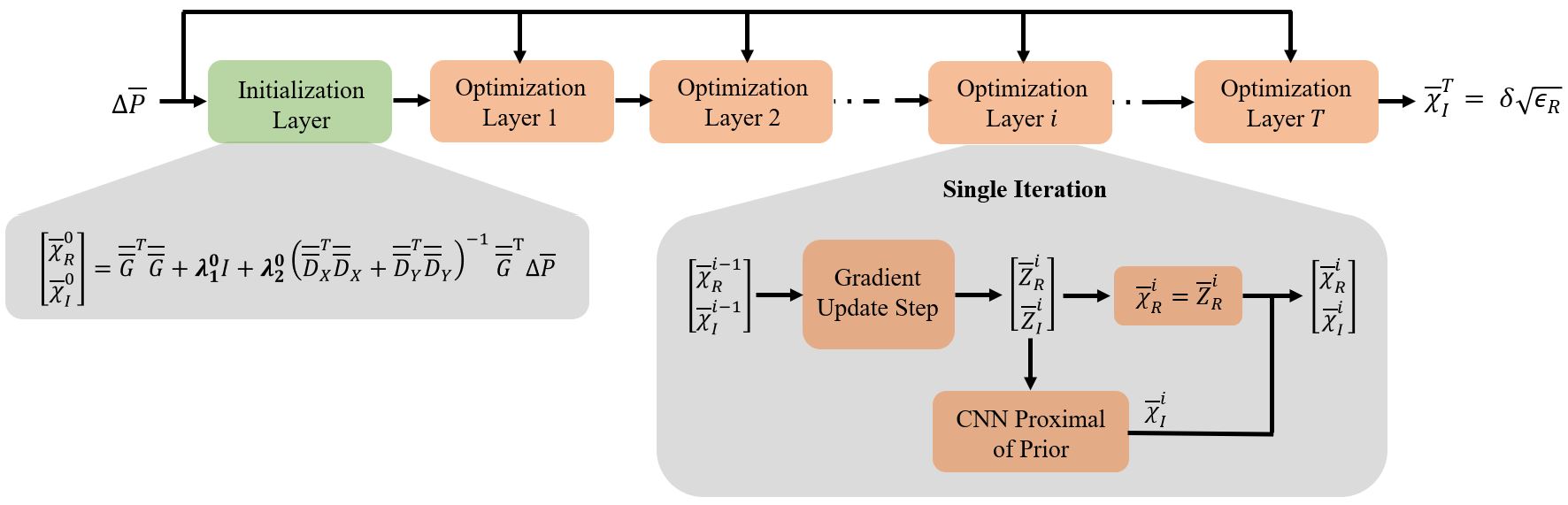}
		\caption{End-to-end unrolled optimization framework}
		\label{Fig_network}
	\end{figure*}
	In this section, we propose an MGDL framework to achieve better estimates of shape and electrical properties ($\delta \sqrt{\epsilon_R}$) of objects inside the DoI by overcoming the following key drawbacks of the xPRA model:
	\begin{enumerate}
		\item Errors due to inversion of the severely ill-posed xPRA inverse problem, and the need to tune regularization parameters during inference,
		\item Errors caused by the model itself, since it is a linear approximation to a non-linear problem, and
		\item Distortions due to the presence of $\theta^s$ and $\theta^i$ terms in the contrast.
	\end{enumerate}
	We show that overcoming these drawbacks can allow us to image objects with very high permittivity, non-convex shapes, and also improve the accuracy of overall shape reconstruction without the need for hyperparameter tuning.

	\subsection{Objective}
	\label{Sec_Objective}
	The goal of our proposed framework is to incorporate the state-of-the-art xPRA model into the deep learning framework to mitigate the aforementioned drawbacks. We do this by combining the xPRA inverse problem objective with traditional and deep learning based priors and solving the resulting optimization problem using an unrolled iterative algorithm.
	
	We start with the model based objective function (\ref{Eq_discrete4}) and reformulate it as a variational regularization problem with multiple penalities on the reconstruction as 
	\begin{equation}
		\label{Eq_objective}
		\begin{aligned}
			&\min_{\chi_R, \chi_I}  \underbrace{\abs*{\abs*{\Delta \overline{P} - \overline{\overline{\mathcal{G}}} 
						\begin{bmatrix}
							\overline{\raisedchi}_R \\
							\overline{\raisedchi}_I
				\end{bmatrix}}}^2_2}_\text{\parbox{1cm}{\centering Data-Fit\\[-4pt] term}}  \\
			&+ \underbrace{\lambda_1 \abs*{\abs*{\overline{\overline{Q}}_1 \left( \chimatrix \right)}}^2_2 
				+ \lambda_2 \abs*{\abs*{\overline{\overline{Q}}_2\left(\chimatrix\right)}}^2_2 }_\text{\parbox{2cm}{\centering Tikhonov\\[-4pt] priors}} 
			+\underbrace{R \left(\chimatrix\right)}_\text{\parbox{1cm}{\centering Deep\\[-4pt] prior}}.
		\end{aligned}		
	\end{equation}
	where $\overline{\overline{Q}}_1$ and $\overline{\overline{Q}}_2$ represent the Tikhonov prior matrices of the classical multi-parameter regularization and $R$ represents the data-driven regularizer which we model as a deep prior. The need for each of these priors is explained in the subsections that follow.
	
	\subsection{Pre-conditioning with Multi-parameter Tikhonov Regularization}
	\label{Sec_MultiReg}
	Unlike inverse problems in other domains such as medical imaging, PD-ISPs are highly ill-posed due to a scarcity of measurements, missing phase information, and the highly non-linear underlying model. This can lead to large errors in the inversion of the poorly conditioned matrix $\overline{\overline{\mathcal{G}}} \in \mathbb{R}^{L\times 2N}$, where $L\ll N$. Hence, a least squares solution provides extremely distorted reconstructions that are highly sensitive to changes in the measurements. Therefore, we add classical Tikhonov priors to pre-condition this highly ill-posed problem and stabilize the inversion.
	
	For our proposed framework, we add two Tikhonov priors to our PD-ISP inverse problem as shown in (\ref{Eq_objective}). $\overline{\overline{Q}}_1$ here represents the identity matrix and the resulting term minimizes the $l_2$ norm of the coefficients in $[\ \overline{\raisedchi}^T_R \ \ \overline{\raisedchi}^T_I\ ]^T$ which is also called Ridge regression. $\overline{\overline{Q}}_2$ represents the difference matrix approximating the derivative operator as the Tikhonov matrix and the resulting term minimizes the $l_2$ norm of the difference between successive coefficient terms (also called H1 regularization). Since we obtain reconstructions of a 2D DoI, we apply the derivative operator in both dimensions. 
	
	The problem in (\ref{Eq_objective}) can now be written as 
	\begin{equation}
		\label{Eq_finalobjective}
		\begin{aligned}
			&\min_{\chi_R, \chi_I} \abs*{\abs*{\Delta \overline{P} - \overline{\overline{\mathcal{G}}} 
					\begin{bmatrix}
						\overline{\raisedchi}_R \\
						\overline{\raisedchi}_I
			\end{bmatrix}}}^2_2
			+\lambda_1 \abs*{\abs*{\chimatrix}}^2_2 \\
			& + \lambda_2 \left( \abs*{\abs*{\overline{\overline{D}}_x\ \chimatrix}}^2_2 + \abs*{\abs*{\overline{\overline{D}}_y\ \chimatrix}}^2_2 \right) 
			+ R \left(\chimatrix \right),
		\end{aligned}
	\end{equation}
	where $\overline{\overline{D}}_x \in \mathbb{R}^{2N\times 2N}$ and $\overline{\overline{D}}_y \in \mathbb{R}^{2N\times 2N}$ are the derivative operators in the horizontal and vertical directions respectively.
	
	The Tikhonov prior terms in (\ref{Eq_finalobjective}) enforce both sparsity and smoothness in the solutions obtained, thus not only estimating important pixels in the solutions but also capturing the spatial continuity in the object of interest. This is especially desirable when we consider large objects with a piece-wise homogeneous distribution of permittivity, which are commonly found in indoor regions. This idea of the prior enforcing both sparsity and smoothness is inspired by Fused Lasso \cite{tibshirani2005sparsity} which is combination of LASSO and Total variation (TV) regularization where an $l_1$ norm penalty is used on both the solution coefficients and their successive differences. However, we restrict ourselves to the use of Tikhonov priors due to their convexity and smoothness properties which can be leveraged to obtain a better convergence rate for algorithms used to solve the optimization problem. Also, the existence of analytical solutions for these priors is useful for initializing the framework, as will be explained in further sections.
	
	\subsection{Optimization Framework}
	\label{Sec_UnrolledOptimizaiton}
	We can rewrite
	(\ref{Eq_finalobjective}) in a compact form as
	\begin{equation}
		\label{Eq_proxobjective}
		\begin{aligned}
			&\min_{\chi_R, \chi_I} f(\overline{\raisedchi}_R, \overline{\raisedchi}_I) + R(\overline{\raisedchi}_R, \overline{\raisedchi}_I),
		\end{aligned}
	\end{equation}
	where $f(\overline{\raisedchi}_R, \overline{\raisedchi}_I)$ is a smooth and differentiable function consisting of the datafit and Tikhonov prior terms and $R(\overline{\raisedchi}_R, \overline{\raisedchi}_I)$ is the deep prior term modeled using CNNs and is therefore non-convex and non-differentiable. This results in (\ref{Eq_proxobjective}) being a non-differentiable objective function, making gradient based methods such as gradient descent or Newton method unsuitable. A better approach is to use Proximal algorithms such as Proximal Gradient Method \cite{parikh2014proximal} or Alternating Direction Method of Multipliers (ADMM) \cite{boyd2011distributed} designed for a non-smooth objective where a proximal step replaces the gradient step for the non-smooth terms. For our proposed framework we use the Proximal Gradient Method (PGM) since it is usually considered to be a good default choice. However, the use of other proximal algorithms can also be explored.
	
	% Explanation for PGM - can remove this to save space later if needed.
	A single step of the Proximal Gradient Method (PGM) for the objective in (\ref{Eq_proxobjective}) can be given as
	\begin{equation}
		\label{Eq_proxgrad}
		\begin{aligned}
			z^{i+1} &= x^i - \eta^i \nabla f(x^i), \\
			x^{i+1} &= \mathbf{prox}_{\eta^i R}(z^{i+1}), \\
			\text{where}\ \mathbf{prox}_{\eta^i R}(u) &= \argmin_x \ R(x) + \frac{1}{2 \eta^i} ||x - u||^2_2,
		\end{aligned}
	\end{equation}
	$\eta^i$ is the step size at the $i^{\text{th}}$ iteration of PGM, $x^i = \left[\overline{\raisedchi}^i_R \ \ \overline{\raisedchi}^i_I \right]^T$ is the input at $i^{\text{th}}$ iteration, $z^{i+1}$ is the gradient step on the smooth function $f$ and $x^{i+1}$ is the output of the proximal operator of $R$ applied to $z^{i+1}$. Here, since $R$ represents the data-driven regularizer in our framework, we directly parameterize the proximal operator of $R$ in the form of a deep CNN, defining the actual regularizer $R$ implicitly in the process.
	
	It is important to note again that our goal is to estimate the $\delta \sqrt{\epsilon_R}$ profile of the DoI and as mentioned in the previous section, this reconstruction ratio is more closely related to $\overline{\raisedchi}_I$ than to $\overline{\raisedchi}_R$. Therefore, we consider $\overline{\raisedchi}_I$ to be the primary variable of our optimization framework from which we can learn the $\delta(\bm{r}_n) \sqrt{\epsilon_R(\bm{r}_n)}$ values of objects, while also using $\overline{\raisedchi}_R$ as an aiding variable to improve it since it provides a good reconstruction of the scatterer boundaries \cite{dubey2022xPRA}.
	
	We now incorporate this discussion mathematically into the optimization algorithm in (\ref{Eq_proxgrad}). We first obtain an initial guess $ \left[\overline{\raisedchi}^0_R \ \ \overline{\raisedchi}^0_I\right]^T$ from the measurements $\Delta \overline{P}$ by analytically solving an objective containing only the datafit and Tikhonov prior terms to obtain
	\begin{equation}
		\label{Eq_initguess}
		\begin{aligned}
			\begin{bmatrix}
				\overline{\raisedchi}^0_R \\
				\overline{\raisedchi}^0_I
			\end{bmatrix} 
			= \biggl(& \overline{\overline{\mathcal{G}}}^T  \overline{\overline{\mathcal{G}}} + \lambda_1^0 I + 
			\lambda_2^0 (\overline{\overline{D}}_x^T \overline{\overline{D}}_x + \overline{\overline{D}}_y^T \overline{\overline{D}}_y)\biggr)^{-1}\  \overline{\overline{\mathcal{G}}}^T \Delta \overline{P} \\
			= & \ \Pi_{\lambda_1^0, \lambda_2^0}  \ \Delta \overline{P},
		\end{aligned}
	\end{equation}
	where $\Pi$ is the regularized pseudo-inverse matrix parameterized by two hyperparameters $\lambda_1^0$ and $\lambda_2^0$, controlling sparsity and smoothness in the initial reconstructions respectively. The values of $\lambda_1^0$ and $\lambda_2^0$ are learned as a part of the training process.  With this initialization, we formulate the PGM algorithm in a way that we can optimize $\overline{\chi}_I$ to bring it close to ground truth $\delta(\bm{r}_n) \sqrt{\epsilon_R(\bm{r}_n)}$ values, while also using information from the aiding variable $\overline{\chi}_R$. To this end, we write the $i^\text{th}$ iteration of the PGM algorithm as
	\begin{subequations}
		\label{Eq_algorithm}
		\begin{align}
			&\begin{bmatrix}
				\overline{Z}_{R}^{i} \\
				\overline{Z}_I^{i}
			\end{bmatrix} = \begin{bmatrix}
				\overline{\chi}_R^{i-1} \\
				\overline{\chi}_I^{i-1}
			\end{bmatrix} - \eta^i \nabla f\left(\begin{bmatrix}
				\overline{\chi}_R^{i-1} \\
				\overline{\chi}_I^{i-1}
			\end{bmatrix} \right), \\
			%		&\chimatrix^{t+1} = \begin{bmatrix}
				%			\overline{Z}_R^{t+1} \\
				%			\mathbf{prox}_{\eta^i R}(\overline{Z}_R^{t+1}, \overline{Z}_I^{t+1})
				%		\end{bmatrix} \\
			&\overline{\raisedchi}^{i}_R = \overline{Z}_R^{i}, \\
			&\overline{\raisedchi}^{i}_I = \mathbf{prox}_{\eta^i R}(\overline{Z}_R^{i}, \overline{Z}_I^{i}),
			%	&\overline{\raisedchi}^{i+1} =  [{\overline{\raisedchi}^{i+1}_R}^T\ \ {\overline{\raisedchi}^{i+1}_I}^T]^T
		\end{align}
	\end{subequations}
	where the gradient of $f$ is obtained from (\ref{Eq_finalobjective}) as
	\begin{equation}
		\begin{aligned}
			\nabla f\left(\begin{bmatrix}
				\overline{\chi}_R^{i-1} \\
				\overline{\chi}_I^{i-1}
			\end{bmatrix} \right)&  = \overline{\overline{\mathcal{G}}}^T \left(\overline{\overline{\mathcal{G}}} \begin{bmatrix}
				\overline{\chi}_R^{i-1} \\
				\overline{\chi}_I^{i-1}
			\end{bmatrix}  - \Delta \overline{P}\right) \\
			&+ \left(\lambda_1^i I + \lambda_2^i (\overline{\overline{D}}_x^T \overline{\overline{D}}_x + \overline{\overline{D}}_y^T \overline{\overline{D}}_y) \right) \begin{bmatrix}
				\overline{\chi}_R^{i-1} \\
				\overline{\chi}_I^{i-1}
			\end{bmatrix}.
		\end{aligned}
	\end{equation}
	Here, $\lambda_1^i$ and $\lambda_2^i$ are the regularization parameters for each iteration of PGM and are learned during training similar to $\lambda_1^0$ and $\lambda_2^0$. Note that $\lambda_1^i$ and $\lambda_2^i$ can be different for different iterations depending on the residual error between the ground truth $\delta\sqrt{\epsilon_R}$ profile and the input $\overline{\chi}_I^i$ for that iteration. Having this flexibility in regularization parameter values for different iterations is often ignored, but it is crucial since as the output improves with iterations, the regularization parameter needs to change.
	
	$\overline{Z}^{i}_R$ and $\overline{Z}^{i}_I$ in (\ref{Eq_algorithm}a) are outputs of the gradient update step corresponding to $\overline{\raisedchi}^{i-1}_R$ and $\overline{\raisedchi}^{i-1}_I$ respectively. Both $\overline{Z}^{i}_R$ and $\overline{Z}^{i}_I$ are given as inputs to the $\mathbf{prox}$ operator modeled as a CNN to learn the ground truth $\delta \sqrt{\epsilon_R}$ profile from $\overline{Z}^{i}_I$ while also using the boundary information contained in $\overline{Z}^{i}_R$. However, it is important to note that CNN does not operate on $\overline{Z}^{i}_R$; it is directly passed to the next iteration after the gradient update step as $\overline{\raisedchi}^{i}_R$ and we only receive $\overline{\raisedchi}^{i}_I$ as the CNN output. This is akin to solving the optimization problem
	\begin{equation}
		\begin{aligned}
			\min_{\chi_I}\ \mathbf{prox}_{\eta^i R} \left(\begin{bmatrix}
				\overline{\chi}_R^{i-1} \\
				\overline{\chi}_I^{i-1}
			\end{bmatrix}  - \eta^i \nabla f\left(\begin{bmatrix}
				\overline{\chi}_R^{i-1} \\
				\overline{\chi}_I^{i-1}
			\end{bmatrix} \right)\right).
		\end{aligned}
	\end{equation} 
	The output of the CNN is thus an improved $\overline{\raisedchi}_I^i$ that is closer to the ground truth $\delta \sqrt{\epsilon_R}$ profile. As we increase the number of iterations, the output of the $\mathbf{prox}$ operator becomes closer and closer to the ground truth $\delta(\bm{r}_n) \sqrt{\epsilon_R(\bm{r}_n)}$ values, eventually giving us accurate reconstructions of the $\delta \sqrt{\epsilon_R}$ profile in the last iteration.
	
	\subsection{End-to-end Unrolled Optimization algorithm}
	\label{Sec_E2E}
	
	To construct an end-to-end reconstruction framework, we first choose a fixed number of iterations $T$ for (\ref{Eq_algorithm}) and unroll the resulting iterative algorithm into a network, with each layer representing one iteration \cite{monga2021algorithm}. This is illustrated in Fig. \ref{Fig_network}. Each layer of this unrolled network consists of a gradient step that is a function of the measurement vector $\Delta \overline{P}$, the xPRA matrix $\overline{\overline{\mathcal{G}}}$, and the Tikhonov prior terms, followed by a $\mathbf{prox}$ operator of the data-driven regularizer parameterized by a CNN. Passing through this network is equivalent to executing the iterative algorithm in (\ref{Eq_algorithm}) $T$ times. Also, algorithm parameters such as the step size $\eta$, the Tikhonov regularization parameters ($\lambda_1^i, \lambda_2^i$) and the deep prior network parameters all transfer to the unrolled network parameters. This unrolled network can then be trained to learn the optimal values of these parameters using training data. Thus, the trained network can be interpreted as a parameter optimized iterative algorithm. Another advantage of unrolled iterative algorithms is that they converge in much fewer iterations as compared to their traditional counterparts \cite{monga2021algorithm}. 
	
	The final MGDL framework is described in Fig. \ref{Fig_network}. The key components of the framework are described below:
	
	\subsubsection{\textbf{Initialization layer}}
	The network obtained by unrolling the iterative algorithm in (\ref{Eq_algorithm}) is preceded by an initialization layer which takes as input the measurement vector $\Delta \overline{P}$ and generates a initial reconstruction $ \left[\overline{\raisedchi}^0_R \ \ \overline{\raisedchi}^0_I\right]^T$ using (\ref{Eq_initguess}). This regularized solution lets us initialize the unrolled network with a noisy but stable reconstruction as opposed to initializing it with random values or the least square solution of a highly ill-posed system, thus leading to faster convergence of the unrolled network. We add a rectified linear unit (ReLU) activation function at the end of this layer to prevent negative $\delta(\bm{r}_n) \sqrt{\epsilon_R(\bm{r}_n)}$ values, since they are not physically possible. Also, the regularization parameters $\lambda_1^0$ and $\lambda_2^0$ do not need to be tuned manually and are learned from data.
	
	%\vspace{0.5em}
	
	\subsubsection{\textbf{Deep Prior}}
	For the proximal operator of $R$ parameterized by a deep network, we use the U-Net architecture \cite{unetoriginal}. An important feature of this architecture is its encoder-decoder structure that leads to a bottleneck in the network, which helps it learn structure in data that cannot be captured by traditional CNNs \cite{deepimageprior, deshmukh2022physics}.
	
	The U-Net prior in each optimization layer contains 4 encoder layers and 4 decoder layers, each of which contain $64$ $3 \times 3$ convolutional filters followed by batch normalization and ReLU activation. It also contains skip connections from the encoder layers to decoder layers which lead to faster training of the network. Specifically, we add skip connections between each encoder layer $i$ and decoder layer $n-i$ where $n$ is the total number of layers in the encoder or decoder. Each skip connection concatenates all channels at the output of decoder layer $n-i$ with those at the output of encoder layer $i$.
	
	In each optimization layer (described in (\ref{Eq_algorithm}) and in Fig. \ref{Fig_network}), the outputs $\overline{Z}^i_R \in \mathbb{R}^{N \times 1}$ and $\overline{Z}^i_I \in \mathbb{R}^{N \times 1}$ of the gradient update step in (\ref{Eq_algorithm}a) are converted to 2D images of dimensions $n_x \times n_y$ to map them to the 2D DoI in Fig. \ref{RTInetwork2}. These 2D images ($\overline{\overline{Z}}^i_R, \overline{\overline{Z}}^i_I \in \mathbb{R}^{n_x \times n_y}$) are then given as inputs to the U-Net as two channels. The output of the
	U-Net is a 2D image representing $\overline{\overline{\raisedchi}}^{i}_I \in \mathbb{R}^{n_x \times n_y}$. This is unrolled into a vector $\overline{\raisedchi}^{i}_I \in \mathbb{R}^{N \times 1}$, concatenated to $\overline{\raisedchi}^{i}_R$ and propagated to the next layer to repeat same process. 
	
	In addition to using conventional regularization in the form of Tikhonov priors that enforce sparsity and smoothness, using data-driven regularization in the form of a deep prior such as U-Net in each layer is useful for learning prior information such as shapes and contrast values of the reconstruction while also learning to remove the distortion caused in $\overline{\raisedchi}_I$ by the $\theta^i$ term. The deep prior also extracts useful information from $\overline{\raisedchi}_R$ to aid the reconstruction. As we move through the optimization layers, $\overline{\raisedchi}_I$ becomes closer and closer to the $\delta \sqrt{\epsilon_R}$ profile, and the U-Net prior in the final optimization layer provides reconstruction of $\delta \sqrt{\epsilon_R}$ profile of the DoI as the output.
	
	\subsubsection{\textbf{Learning Hyper-parameters and Network Parameters}}
	All hyperparameters involved in the optimization process including the regularization parameters in the initialization layer $(\lambda_1^0, \lambda_2^0 )$ in (\ref{Eq_initguess}) and the regularization parameters $(\lambda_1^i, \lambda_2^i)$ and the learning rates $(\eta^i)$ in the $i^{th}$ optimization layer in (\ref{Eq_algorithm}) are learned during the training of the unrolled network. Also, as mentioned before, $\lambda_1^i$ and $\lambda_2^i$ can vary across layers depending on the residual error between the ground truth $\delta\sqrt{\epsilon_R}$ profile and $\overline{\chi}_I^i$. This leads to different intensities of regularization being applied at different iterations/layers, and as the reconstruction gets better in later layers, the regularization parameter can adapt to the residual noise and distortion. Also, even though the same U-Net architecture is used as the deep prior in all optimization layers, we parameterize them as separate CNNs. This offers tremendous flexibility by allowing the prior to learn a specialized function at every layer based on the noise and distortions in reconstructions in that layer. 
	
	To build our end-to-end reconstruction framework, we unroll the PGM algorithm for $T=5$ iterations and precede it by the initialization layer. This complete network is then trained using a mean squared error loss and Adam optimizer. The key issue in training this framework is that the scale at which regularization parameters vary can be very different from the scale at which other network parameters such as the U-Net weights change. Regularization parameters for a layer can typically vary on a logarithmic scale and take different values depending on distortions in reconstructions in that layer. Therefore, we cannot use an optimizer with the same learning rate for both these types of parameters. To tackle this, we use a learning rate of $10^{-2}$ for the regularization parameters in all layers ($\lambda_1^0, \lambda_2^0, \lambda_1^i, \lambda_2^i$) and $10^{-4}$ for all other parameters throughout the network. The higher learning rate of regularization parameters offers greater flexibility in tuning these parameter values correctly since very small changes to these do not have a significant effect on the reconstructions.
	
	\section{Simulation Results}
	\label{Sec_SimulationResults}
	This section provides simulation examples to evaluate the performance of our proposed framework. The setup used for the simulations is compatible with that used for the experiments (experimental results are presented in the next section). In the simulations, the values of the size and relative permittivity $\epsilon_r = \epsilon_R+j \epsilon_I$ of scatterers are selected such that these represent objects found in a typical indoor environment. The PSNR results for each reconstruction are also provided for quantitative evaluation.
	
	\subsection{Simulation Setup}
	\label{Sec_ResultSetup}
\begin{figure}[h]
		\centering
		\begin{subfigure}{0.21\textwidth}
			\centering
			\includegraphics[width=\textwidth]{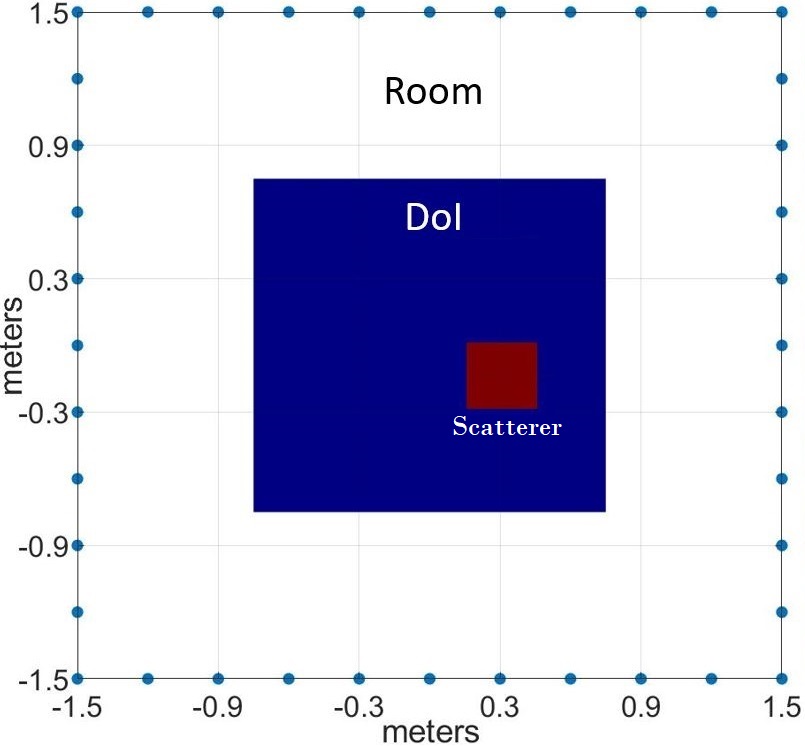}
			\subcaption{Numerical Setup}
		\end{subfigure}
		\begin{subfigure}{0.26\textwidth}
			\centering
			\includegraphics[width=\textwidth]{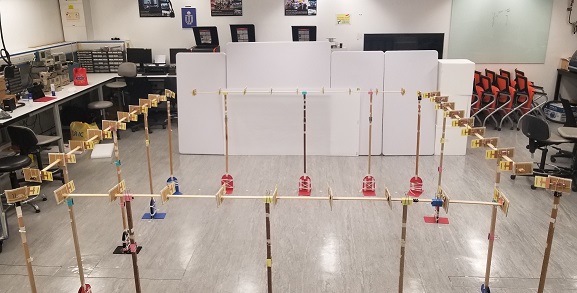}
			\subcaption{Experimental Setup}
		\end{subfigure}
		\caption{Data Acquisition Setup. (a) Simulation setup where transceiver nodes (shown as blue dots) are placed at boundary of a $3\times 3$ m$^2$ room. The $1.5\times 1.5$ m$^2$ DoI is inside this room, shown as blue colored region. (b) Experimental setup with same configuration as simulation setup in (a).}
		\label{problemsetup}
	\end{figure}

	We consider a $3\times3$ m$^2$ indoor region for both our simulation and experiments. It contains a $1.5\times1.5$ m$^2$ DoI to be imaged as shown in Fig. \ref{problemsetup} (a). We simulate transceiver nodes operating at 2.4 GHz with incident wavelength $\lambda_0 = 12.5$ cm (in the experiments these are Wi-Fi transceivers). The size of the DoI is $12 \lambda_0 \times 12 \lambda_0$ and there are $40$ equidistant transceivers $(M = 40)$ placed at the DoI boundary to collect measurements. Therefore, the total number of wireless links is $L = M(M-1)/2 = 780$. We discretize the DoI into $160000$ grids $(400 \times 400)$ for the forward problem and $N=2500$ grids $(50 \times 50)$ for the inverse problem. To generate the forward measurement data (RSS values) for training, we use the Method of Moments approach detailed in \cite{chen2018computational}.
	
	\subsection{Training Data}
	\label{Sec_TrainingData}
	For training the network, we generate a dataset containing circular and square shaped objects placed in the DoI (see sample object profiles in Fig. \ref{Results_ex1}). A single object can be a circle or a square with  equal probability. Since the U-Net model used as a deep prior is translationally invariant, we only place objects in the upper half of the DoI. Therefore, assuming that the origin lies at the center of the DoI, the $x$ and $y$ co-ordinates of the object centers are sampled uniformly from $c_x=[-0.6, 0.6]$ and $c_y=[0.15, 0.6]$ respectively.

	Since we are demonstrating the effectiveness of our framework for indoor imaging, we select the real component of the relative permittivity of objects $\epsilon_R$ based on different objects and material typically found in indoor regions such as wood, concrete walls, furniture, etc. At $2.4$ GHz, these objects typically have $\epsilon_R$ values ranging from $2$ to $10$ \cite{4562803, ahmad2014partially, Productnote}. Also, various organs of the human body can have permittivity values ranging from $50$ to $77$ because of the high water content (water at 2.4 GHz has $\epsilon_R = 77$ \cite{4562803, ahmad2014partially, Productnote}). Therefore, for generating training data, we sample $\epsilon_R$ uniformly at intervals of $2$ in the ranges $2 \leq \epsilon_R \leq 10$ and $50 \leq \epsilon_R \leq 77$. For each of these permittivity values, we use a loss tangent of $\delta = 0.1$ such that the imaginary part of permittivity $\epsilon_I = \delta \epsilon_R$, therefore the objects considered are low-loss. This is consistent with typical objects in indoor environments \cite{4562803, ahmad2014partially, Productnote}. The resulting object profiles with $\epsilon_r(\bm{r}_n) = \epsilon_R(\bm{r}_n) + j\epsilon_I(\bm{r}_n)$ in the DoI are then used to generate the forward RSS measurement data as explained in the previous section. To solve the inverse problem, we obtain the $\delta(\bm{r}_n) \sqrt{\epsilon_R(\bm{r}_n)}$ values for each object. This forms the ground truth image which is used in addition to the measurement data to train the end-to-end framework.
	
	The size of the objects (diameter for circular objects and length of side for square objects) are also sampled uniformly from the set $S_{L} \in \{0.5\lambda_0, \lambda_0, 1.5\lambda_0, 2\lambda_0, 2.5\lambda_0\}$. Therefore, the largest single object in the dataset has a size of 2.5 $\lambda_0$ with $\epsilon_R = 77$, making its electrical size $\approx 20\lambda$ where $\lambda = \lambda_0/\sqrt{77}$, which is extremely large.
	
	Since we consider permittivity values as high as $\epsilon_R = 77$, multiple scattering effects among objects are expected to be significant. These effects cannot be predicted by the framework if it is only trained on samples containing a single object. Therefore, each of the training samples we generate contain two objects in the DoI. As shown in the following sections, this can easily generalize to other settings.
	
	We generate 2000 such samples, and divide them into training, validation and test datasets, using 1350 samples for training, 150 samples for validation and the remaining 500 samples for testing the trained framework for generalization to samples not present in the training data. We call this test data the \textit{in-sample} test data, since the objects in these samples have the same parameters as the ones used in training.
	
	To better test the proposed framework, we also generate an \textit{out-of-sample} test data in which object parameters differ from the ones mentioned in this section, i.e. objects have shapes other than circles and squares and have permittivity values different from the ones used in training. The performance of our proposed framework on this out-of-sample dataset is also demonstrated in this section.

	For brevity, from here on, we refer to our proposed MGDL framework using the xPRA model, Tikhnov priors and the deep prior as \textbf{xPRA-TK-DPrior}. 
	
	\subsection{Reconstruction results}
	
	\begin{figure*}[!h]
		\centering
		%	\adjustbox{minipage=0.05\textwidth}{\subcaption*{}}%
		\adjustbox{minipage=0.18\textwidth}{\subcaption*{\textbf{\ \ \ Ground Truth: $\delta \sqrt{\epsilon_R}$}}}%
		\adjustbox{minipage=0.18\textwidth}{\subcaption*{\textbf{\ \ xPRA-TV}}}%
		\adjustbox{minipage=0.18\textwidth}{\subcaption*{\textbf{        DI}}}%
		\adjustbox{minipage=0.18\textwidth}{\subcaption*{\textbf{    xPRA-DPrior}}}%
		\adjustbox{minipage=0.18\textwidth}{\subcaption*{\textbf{xPRA-TK-DPrior}}}%\\
		\vspace{-0.2\baselineskip}
		%%%%%%%%%%%%%%%%%%%%%%%%%%%%		
		\begin{subfigure}{\textwidth}
			\centering
			\includegraphics[scale=0.38]{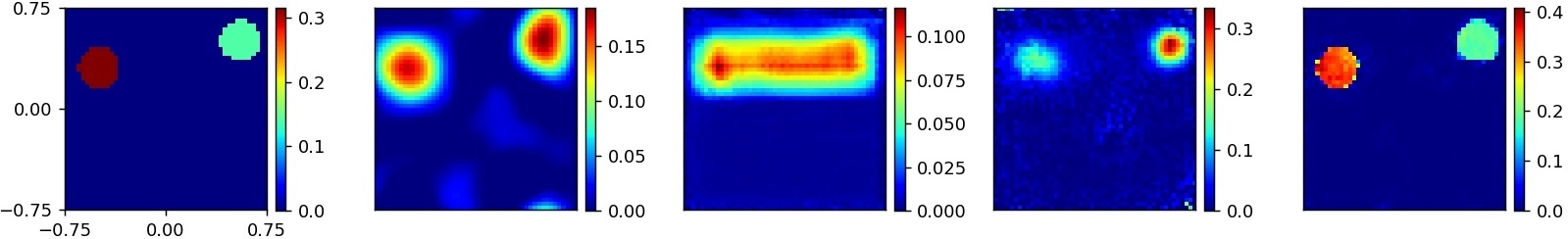}
			\subcaption{$\epsilon_r = 2 + 0.2j\ (\delta \sqrt{\epsilon_R} = 0.141), 10  + 1j\ (\delta \sqrt{\epsilon_R} = 0.316)$}
		\end{subfigure}
		\begin{subfigure}{\textwidth}
			\centering
			\includegraphics[scale=0.475]{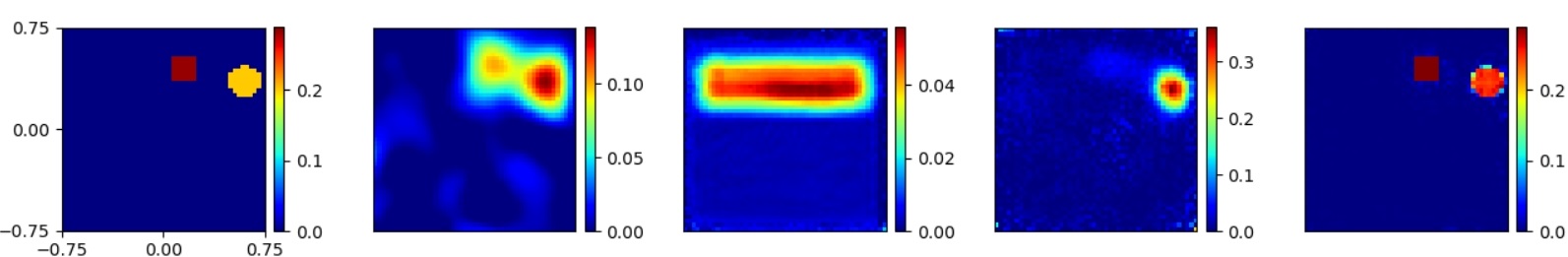}
			\subcaption{$\epsilon_r = 4 + 0.4j\ (\delta \sqrt{\epsilon_R} = 0.2), 8  + 0.8j\ (\delta \sqrt{\epsilon_R} = 0.282)$}
		\end{subfigure}
		\begin{subfigure}{\textwidth}
			\centering
			\includegraphics[scale=0.475]{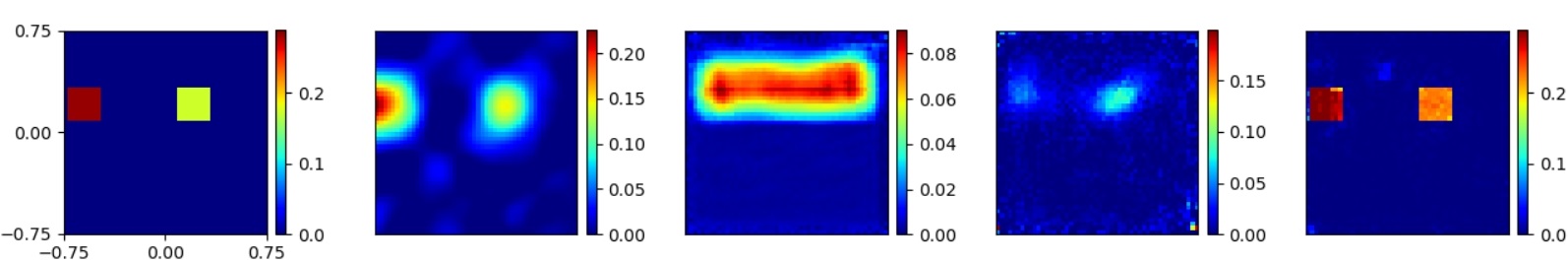}
			\subcaption{$\epsilon_r = 3 + 0.3j\ (\delta \sqrt{\epsilon_R} = 0.173), 8  + 0.8j\ (\delta \sqrt{\epsilon_R} = 0.282)$}
		\end{subfigure}
		\begin{subfigure}{\textwidth}
			\centering
			\includegraphics[scale=0.475]{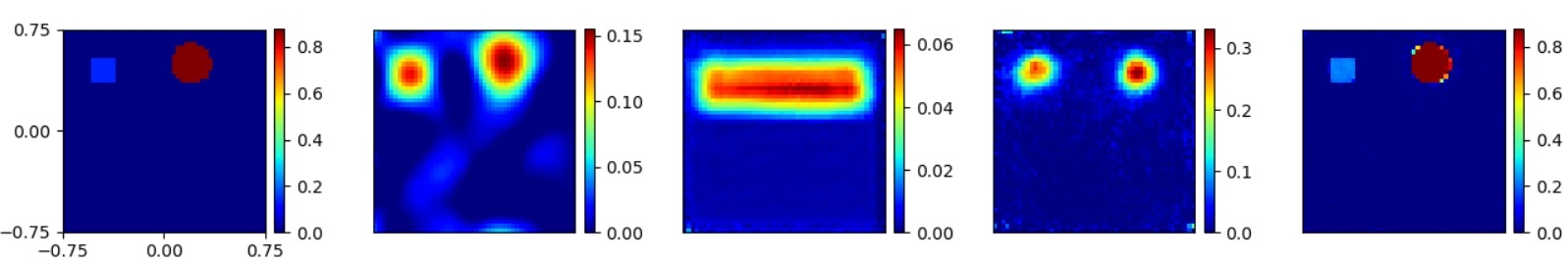}
			\subcaption{$\epsilon_r = 2 + 0.2j\ (\delta \sqrt{\epsilon_R} = 0.141), 77  + 7.7j\ (\delta \sqrt{\epsilon_R} = 0.877)$}
		\end{subfigure}
		\begin{subfigure}{\textwidth}
			\centering
			\includegraphics[scale=0.38]{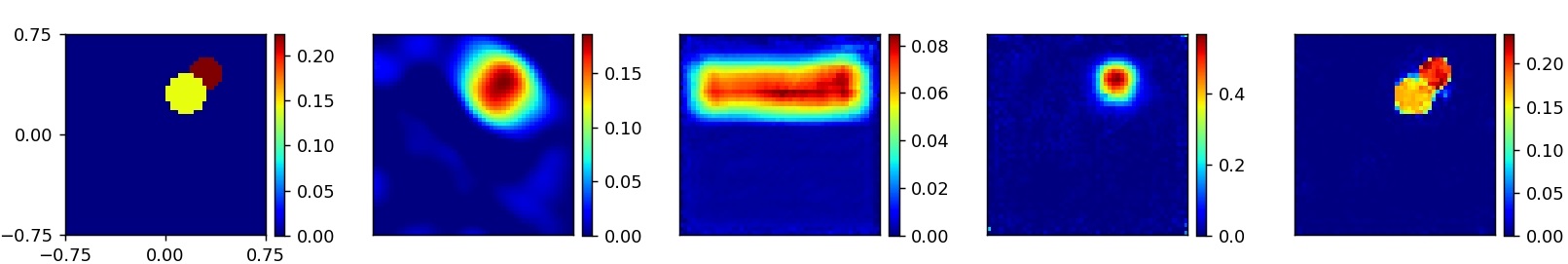}
			\subcaption{$\epsilon_r = 5 + 0.5j\ (\delta \sqrt{\epsilon_R} = 0.223), 2  + 0.2j\ (\delta \sqrt{\epsilon_R} = 0.141)$}
		\end{subfigure}
		\caption{Reconstruction results for in-sample test data where the reconstructed  $\delta(\bm{r}_n) \sqrt{\epsilon_R(\bm{r}_n)}$ values are indicated by color and the exact permittivities of the ground truth objects are listed in the subcaptions. The first column shows the ground truth, and the columns from second to last show reconstructions obtained using xPRA-TV, DI, xPRA-DPrior and xPRA-TK-DPrior respectively. Dimensions are shown on the x- and y-axis of the ground truth results in the first column and are in units of meters. Also note that the color scales in the second, third and fourth columns of images are different from those in the first and final columns.   }
		\label{Results_ex1}
	\end{figure*}
	
	To demonstrate the comparative performance of our proposed xPRA-TK-DPrior framework, we compare it to other state-of-the-art end-to-end frameworks including the unrolled ADMM total variation regularized approach, the direct inversion approach and the existing deep prior approach. Specifically, we compare the proposed framework to the following:
	\begin{enumerate}[leftmargin=0.45cm]
		\item \textbf{xPRA model with Total Variation regularization (xPRA-TV)}: To analyze the effect of the deep prior on the reconstructions obtained, we remove it and instead solve the optimization problem in (\ref{Eq_discrete4}) with a widely-used Total Variation prior as
		\begin{equation}
			\label{Eq_ADMMTV}
			\begin{aligned}
				&\min_{\chi_R, \chi_I} \abs*{\abs*{\Delta \overline{P} - \overline{\overline{\mathcal{G}}} 
						\begin{bmatrix}
							\overline{\raisedchi}_R \\
							\overline{\raisedchi}_I
				\end{bmatrix}}}^2_2 \\
				& + \lambda_{TV} \left( \abs*{\abs*{\overline{\overline{D}}_x\ \chimatrix}}_1 + \abs*{\abs*{\overline{\overline{D}}_y\ \chimatrix}}_1 \right),
			\end{aligned}
		\end{equation}
		where $\overline{\overline{D}}_x \in \mathbb{R}^{2N\times 2N}$ and $\overline{\overline{D}}_y \in \mathbb{R}^{2N\times 2N}$ are the derivative operators in horizontal and vertical directions respectively. We solve (\ref{Eq_ADMMTV}) using the ADMM algorithm which we unroll for $T = 5$ iterations and learn the value of $\lambda_{TV}$ while training the unrolled network, similar to the regularization parameters learned in our framework.

		\item \textbf{Direct Inversion (DI):} For the sake of providing a complete picture, we also observe the effect of excluding the xPRA forward model in our framework. A framework that does not use the xPRA model only consists of the initialization layer followed by a cascade of $T$ deep priors. This is an example of the \textit{Direct Inversion} technique discussed in Section \ref{Sec_Introduction}. In this case the initialization layer consists of a fully connected layer that directly translates measurements $\Delta \overline{P} \in R^{780\times1}$ to the reconstruction domain $ \left[\overline{\raisedchi}^0_R \ \ \overline{\raisedchi}^0_I\right]^T \in R^{5000 \times 1}$.
		
		\item \textbf{xPRA model with only Deep prior (xPRA-DPrior):} In order to understand how much of the reconstruction performance can be attributed to preconditioning the highly ill-conditioned xPRA model matrix using multi-parameter regularization, we also initialize and train the framework by omitting the Tikhonov prior terms. The resulting framework is an xPRA based deep prior technique that is similar to state-of-the-art deep prior based methods used to solve inverse problems in the field of medical imaging and image enhancement \cite{diamond2017unrolled}. We call this xPRA-DPrior. However, as we have mentioned before, since PD-ISPs are much more ill-posed as compared to problems in these domains, pre-conditioning the model matrix becomes important. We can see the importance of this pre-conditioning for PD-ISPs by comparing this framework with our proposed framework that contains additional Tikhonov priors.
		
	\end{enumerate}
	
	The summary and average PSNR values for all these frameworks over the in-sample test data are provided in Table \ref{Table_compare}.
	
	In terms of performance evaluation of xPRA model, our recent work \cite{dubey2022xPRA} is used as a reference since it significantly outperforms the state-of-the-art approach known as subspace based optimization method using phaseless data (PD-SOM) \cite{chen2018computational, xu2020deep, chen2010subspace}. Therefore, we do not provide comparison with PD-SOM in this work.
	
	We train all these frameworks (\textbf{xPRA-TV}, \textbf{DI}, \textbf{xPRA-DPrior} and \textbf{xPRA-TK-DPrior}) for $20$ epochs on a training data of $1500$ samples as explained in Section \ref{Sec_TrainingData}.

	\begin{table}[t]
		\begin{center}
			\begin{tabular}{c c c}
				\hline \hline
				\textbf{Framework} & \textbf{Description} & \textbf{\shortstack{\\Average \\ PSNR (dB)}} \\ [0.5ex] 
				\hline
				\textbf{xPRA-TV} & \shortstack{\\ xPRA with TV regularization \\solved using unrolled ADMM} & 22.12 \\% [1ex] 
				\hline
				\textbf{DI} & \shortstack{\\ Cascade of deep priors without \\ xPRA.} & 21.98 \\ %[1ex] 
				\hline
				\textbf{xPRA-DPrior} & \shortstack{\\ xPRA with deep prior. \\ Solved  using unrolled \\ Proximal Gradient method.} &  22.67 \\% [1ex] 
				\hline
				\textbf{{xPRA-TK-DPrior}} & \shortstack{\\ Our proposed framework: xPRA \\ with Tikhonov and deep priors.\\ Solved using unrolled \\ Proximal Gradient method.} &  \textbf{33.93} \\ %[1ex] 
				\hline \hline
			\end{tabular}
		\end{center}
		\caption{Comparison of different end-to-end frameworks using average PSNR values for reconstructions.}
		\label{Table_compare}
	\end{table}
	
	Fig. \ref{Results_ex1} shows results for samples contained in the in-sample test data. The first column shows the ground truth 	$\delta \sqrt{\epsilon_R}$ profiles and the other columns show reconstructions using xPRA-TV, DI, xPRA-DPrior and xPRA-TK-DPrior frameworks respectively. The complex-valued relative permittivity values for the samples are listed in the respective sub-figure captions. These values of relative permittivity are considered extremely high in the inverse scattering community and can create extremely strong scattering, especially since the object sizes are comparable to or larger than $\lambda_0$. 
	
	It can be seen that for all five test profiles shown in Fig. \ref{Results_ex1}, our proposed xPRA-TK-DPrior framework significantly outperforms all other frameworks. The average PSNR values of reconstructions on the in-sample test data for all these frameworks are listed in Table \ref{Table_compare}. 
	
	The DI technique performs the worst as shown in Fig. \ref{Results_ex1} and Table \ref{Table_compare}. This is because it contains no model information to aid reconstructions and therefore needs to learn the non-linear relationship between the measurements $\Delta\overline{P}$ and the scatterer profile entirely from data. Doing so is a highly data intensive task for which a dataset of $1500$ samples is insufficient. Therefore, the DI framework is only able to localize objects to the upper half of the DoI, but it cannot localize them further without additional training data. Fig. \ref{Results_ex1} shows that unlike the DI approach, xPRA-TV method is able to correctly estimate the location of objects. But it cannot accurately estimate their shape and reconstruction amplitude. This is because TV regularization in xPRA-TV makes it a fully linear method. Therefore, it is not able to correct errors caused by the linear approximation (xPRA) of the underlying non-linear forward model. The xPRA-DPrior framework that contains the xPRA model along with deep priors provides much better results as compared to both the fully linear xPRA-TV and the highly non-linear DI frameworks. It does so on account of the framework using the xPRA model to aid reconstructions and the deep prior handling the ill-posedness and learning the intricate spatial structures caused by the underlying non-linearity in the forward model. However, as we can see in the fourth column of Fig. \ref{Results_ex1}, even this technique does not provide highly accurate reconstructions of shape and  $\delta(\bm{r}_n) \sqrt{\epsilon_R(\bm{r}_n)}$ values of objects. It is also important to note that in the xPRA-DPrior framework, we use a U-Net architecture with skip connections, similar to our proposed xPRA-TK-DPrior framework. However the original work related to deep priors \cite{diamond2017unrolled} uses a much simpler CNN architecture. Therefore, the xPRA-DPrior results in Fig. \ref{Results_ex1} are better than the existing deep prior-based methods due to our choice of the U-Net architecture and xPRA model.
	
	The last column in Fig. \ref{Results_ex1} shows the reconstructions using our proposed xPRA-TK-DPrior framework. Similar to xPRA-DPrior, our framework uses the xPRA model to aid reconstructions and deep priors to learn prior information about reconstructions and correct errors in xPRA due to its linear nature. However, unlike xPRA-DPrior in which the highly ill-conditioned and approximate xPRA model causes high distortions in the initialization and subsequent iterations, pre-conditioning it using the multi-parameter regularization stabilizes the solution obtained and results in highly accurate reconstructions of the shape and  $\delta(\bm{r}_n) \sqrt{\epsilon_R(\bm{r}_n)}$ values of all objects.
	
\begin{figure}[!t]
		\centering
		\includegraphics[scale=0.4]{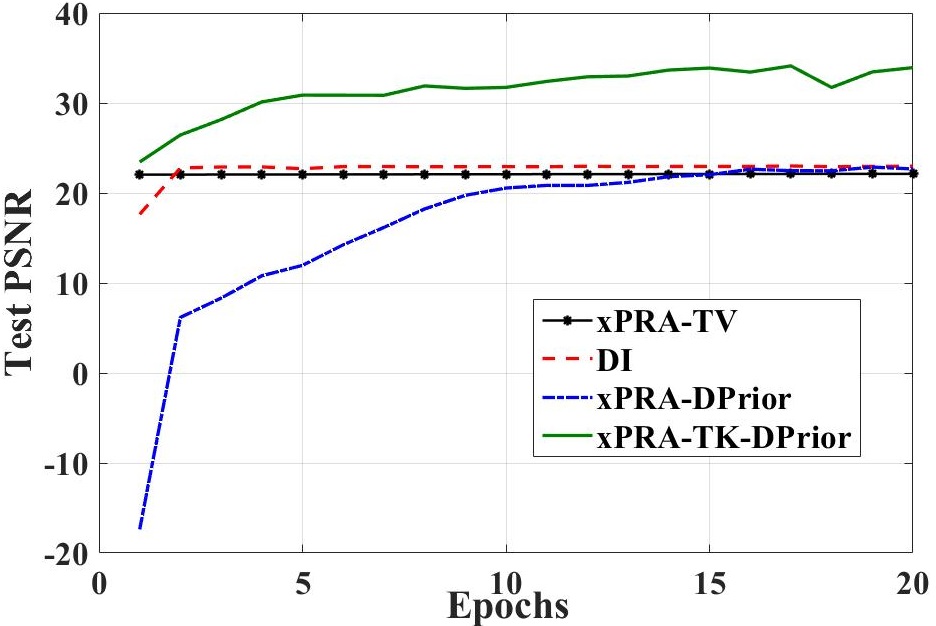}
		\caption{\centering{Performance comparison of different end-to-end frameworks using average PSNR values on test data}}
		\label{psnr_model}
	\end{figure}

	We can also see the superior performance of our proposed framework in Fig. \ref{psnr_model} where it converges to a higher average PSNR value on the in-sample test data, performing much better than all other frameworks in comparison. 
	
	\subsection{Generalization Tests}
\begin{figure}[h]
		\centering
		\begin{subfigure}{0.5\textwidth}
			\centering
			\includegraphics[scale=0.18]{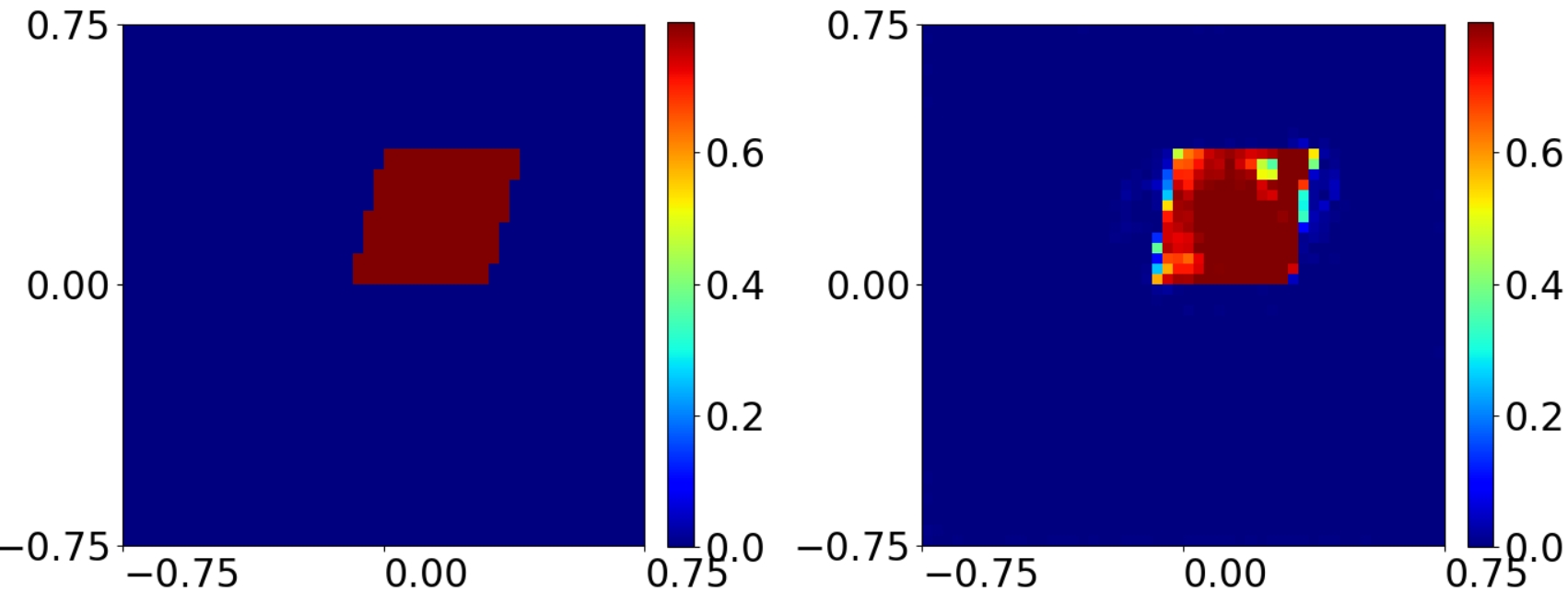}
			\subcaption{}
		\end{subfigure}
		\begin{subfigure}{0.5\textwidth}
			\centering
			\includegraphics[scale=0.18]{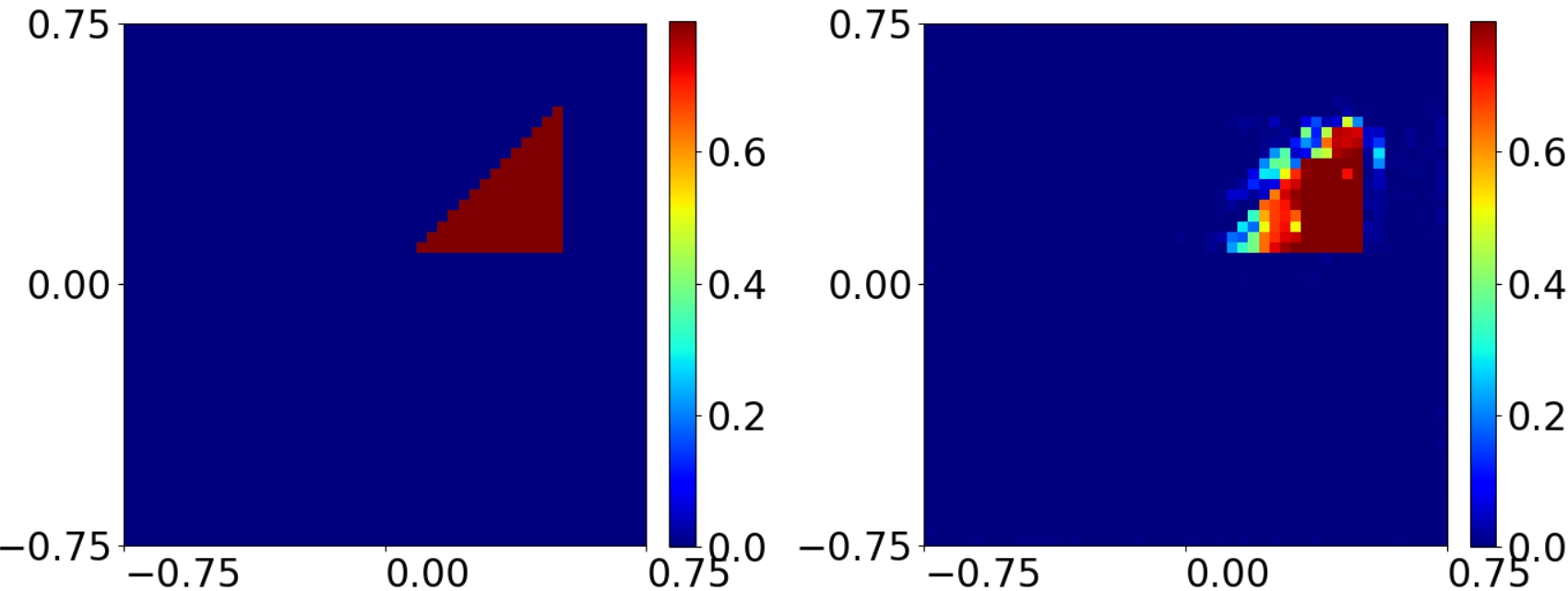}
			\subcaption{}
		\end{subfigure}
		%	\begin{subfigure}{0.5\textwidth}
			%		\centering
			%		\includegraphics[scale=0.22]{results/gen_4.jpg}
			%		\subcaption{20.48 dB}
			%	\end{subfigure}
		%	\begin{subfigure}{0.5\textwidth}
			%		\centering
			%		\includegraphics[scale=0.22]{results/gen_5.jpg}
			%		\subcaption{28.33 dB}
			%	\end{subfigure}
		\caption{Reconstruction results for out-of-sample test data where the reconstructed $\delta(\bm{r}_n) \sqrt{\epsilon_R(\bm{r}_n)}$  values are indicated by color. First column shows the ground truth and the second column shows the reconstruction using xPRA-TK-DPrior. The respective PSNR values of the reconstruction are (a) 26.44 dB, and (b) 25.46 dB. Dimensions shown on the x- and y-axis and are in units of meters.}
		\label{Results_ex2}
	\end{figure}

	The results shown in Fig. \ref{Results_ex1} are for in-sample test data. To test the generalization of our proposed framework, we also evaluate it on out-of-sample test data in which the sample properties differ from those used in training. The results of some of the generalization tests are shown in Fig. \ref{Results_ex2}. Fig. \ref{Results_ex2} (a) and (b) both contain a single object having shapes different from squares or circles, whereas the proposed framework is only trained on profiles that always consist of circular or square shaped scatterers. It can be seen that our proposed framework provides accurate shape and 	$\delta(\bm{r}_n) \sqrt{\epsilon_R(\bm{r}_n)}$ value reconstructions for such samples, showing the high generalization ability of our proposed framework in terms of reconstructing objects with different shapes.
	\begin{figure}[h]
		\centering
		\begin{subfigure}{0.5\textwidth}
			\centering
			\includegraphics[scale=0.18]{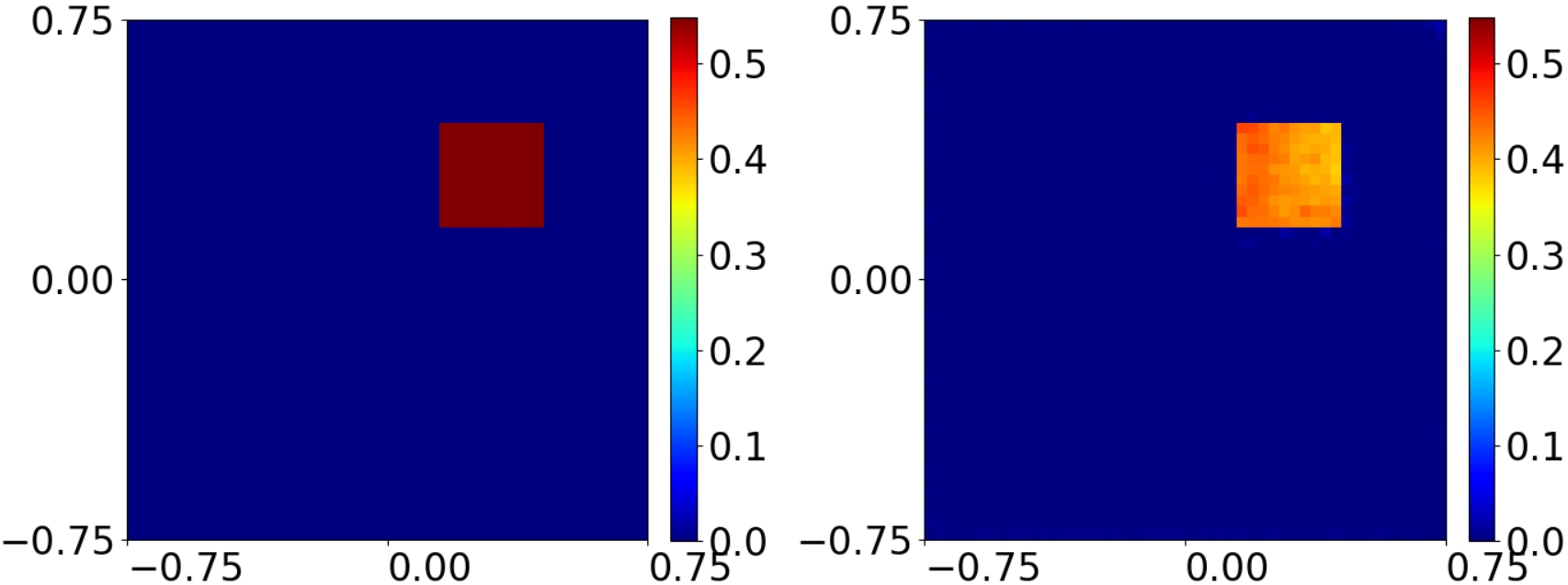}
			\subcaption{$\epsilon_r = 30 + 3j$}
		\end{subfigure}
		\begin{subfigure}{0.5\textwidth}
			\centering
			\includegraphics[scale=0.18]{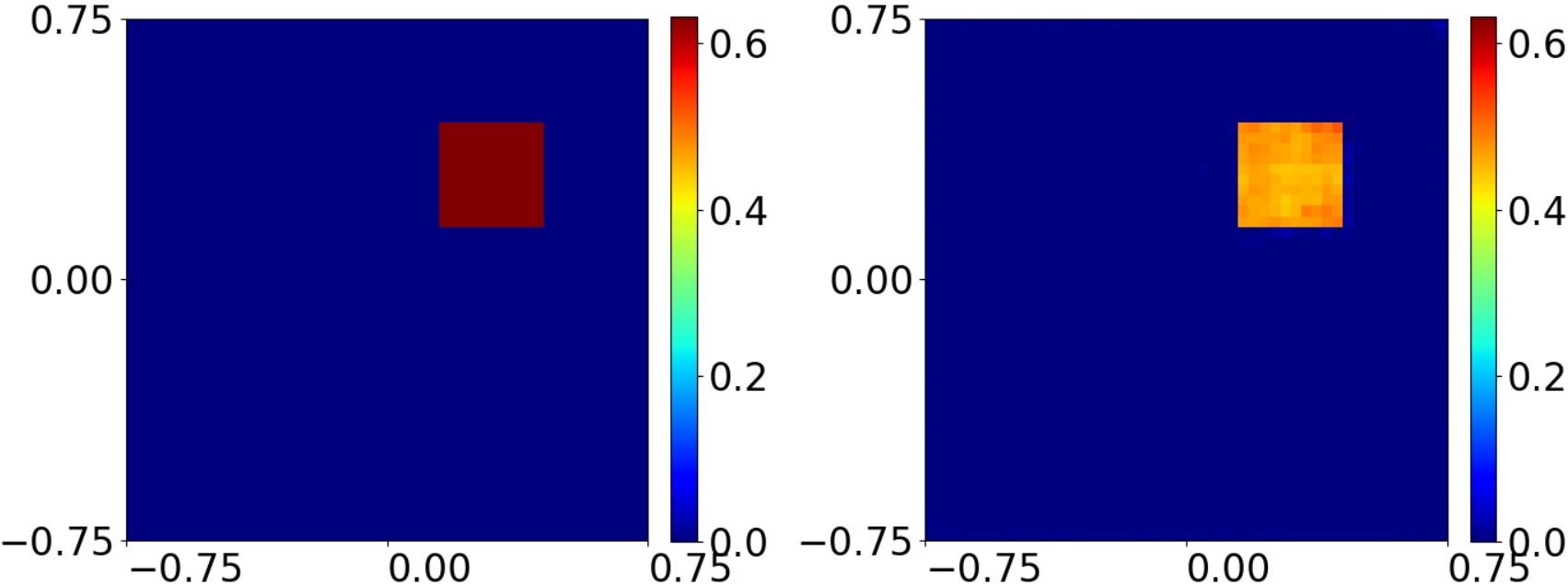}
			\subcaption{$\epsilon_r = 40 + 4j$}
		\end{subfigure}
		%	\begin{subfigure}{0.5\textwidth}
			%		\centering
			%		\includegraphics[scale=0.22]{results/perm_gen_50.jpg}
			%		\subcaption{$\epsilon_r = 50 + 5j$, 30.36 dB}
			%	\end{subfigure}
		%	\begin{subfigure}{0.5\textwidth}
			%		\centering
			%		\includegraphics[scale=0.22]{results/perm_gen_30_77.jpg}
			%		\subcaption{}
			%	\end{subfigure}
		\caption{Reconstruction results for out-of-sample permittivity where the reconstructed $\delta(\bm{r}_n) \sqrt{\epsilon_R(\bm{r}_n)}$ values are indicated by color. First column shows ground truth and the second column shows reconstruction obtained using xPRA-TDPrior. The respective PSNR values of the reconstructions are (a) 31.62 dB, and (b) 29.53 dB. Dimensions shown on the x- and y-axis are in units of meters. }
		\label{Results_ex3}
	\end{figure}
	
	Fig. \ref{Results_ex3} shows generalization tests for $\epsilon_r$ values different from the ones used in training. Fig. \ref{Results_ex3}(a) and (b) show results for a single object that has $\epsilon_r = 30 + 3j$ and $40 + 4j$ respectively. These permittivity values are not present in our training dataset. It can be seen that our proposed framework reconstructs the $\delta \sqrt{\epsilon_R}$ profile for these samples with high accuracy. Therefore, our proposed framework can generalize to both object shapes and reconstruction amplitudes not present in the training data. 
	
	An important point to note for examples in Figs. \ref{Results_ex1}, \ref{Results_ex2} and \ref{Results_ex3} is that all objects considered are piecewise homogeneous in permittivity. This is not a constraint imposed by our framework, but rather by the xPRA model, and is a problem that needs to be explored in future.
	
	\section{Experiment results}
	\label{Sec_ExperimentResults}
	\subsection{Experiment Setup}
	The experiment setup we use for indoor imaging is shown in Fig. \ref{problemsetup} (b) where the environment is approximated as a 2D electromagnetic problem. The DoI is a 1.5$\times$1.5 m$^{2}$ area that lies in a 3$\times$3 m$^{2}$ 2D planar cross-section parallel to the floor. The 1.5$\times$1.5 m$^{2}$ DoI is discretized into grids of 50$\times$50 (2500 grids) for the inverse problem. Similar to the setup described in Section \ref{Sec_ResultSetup}, the experiment also uses $M=40$ Wi-Fi transceiver nodes which are placed at the edge of the 2D cross-section at a height of $1.2$ m above the floor, thus creating a total of $L = M(M-1)/2 = 780$ measurement links. Each node consists of a SparkFun ESP32 Thing board with an integrated Wi-Fi transceiver operating at 2.4 GHz, with the inbuilt omni-directional antennas of the SparkFun ESP32 boards replaced by directive Yagi antennas of 6.6 dBi. More information about the setup can be found in \cite{dubey2021enhanced}. 
	
	\subsection{Experiment Tests}
	\label{Sec_ExperimentTests}
	To obtain reconstructions for measurement data obtained from experiments, we use the same end-to-end network of our proposed framework that is trained on the simulation data described in Section \ref{Sec_TrainingData}. We test our proposed framework on measurement data obtained using some of the objects that can be commonly found in an indoor region.
	
	\begin{figure}[t]
		\centering
		\begin{subfigure}{0.5\textwidth}
			\centering
			\includegraphics[scale=0.075]{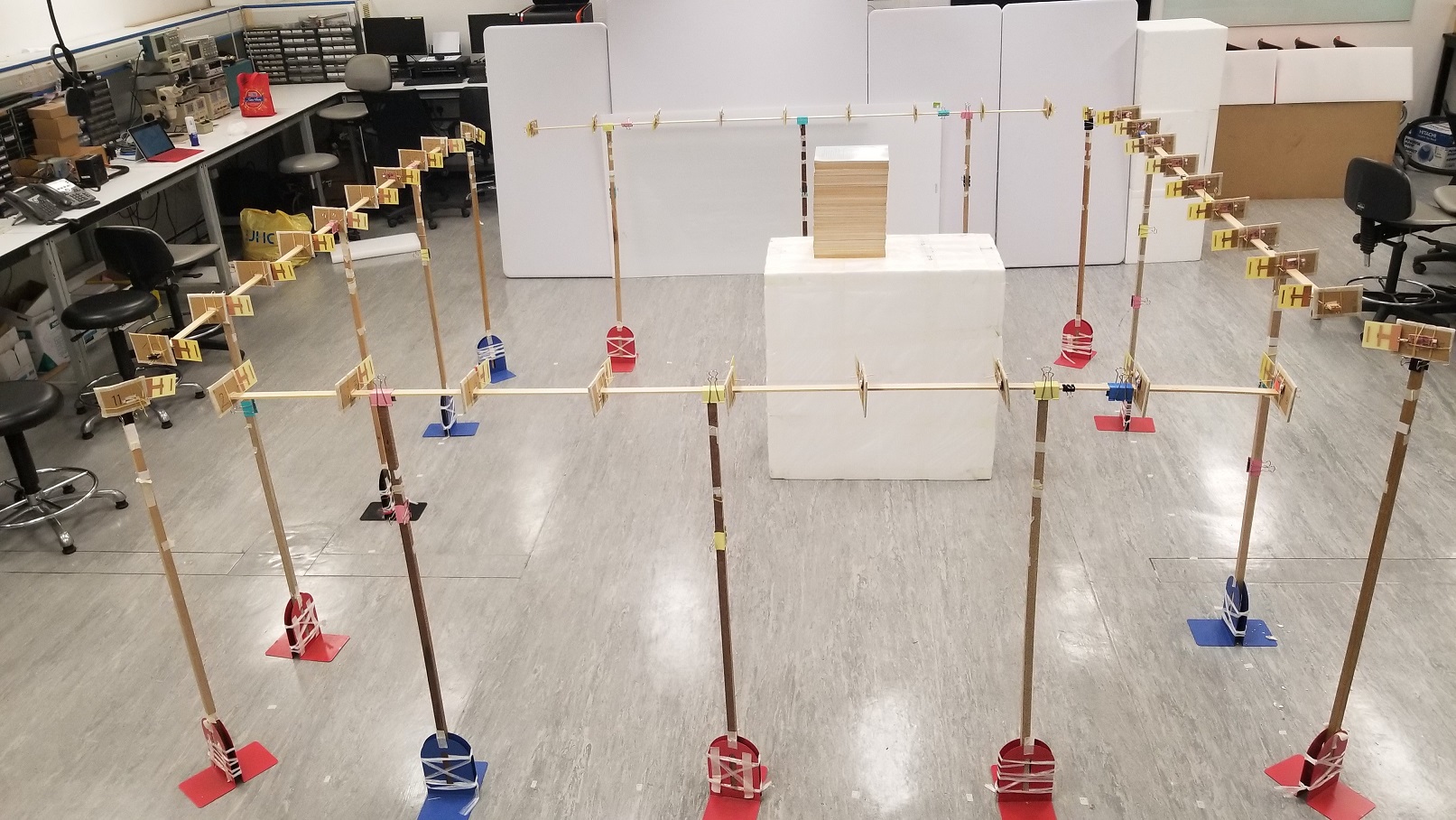}
			\includegraphics[scale=0.18]{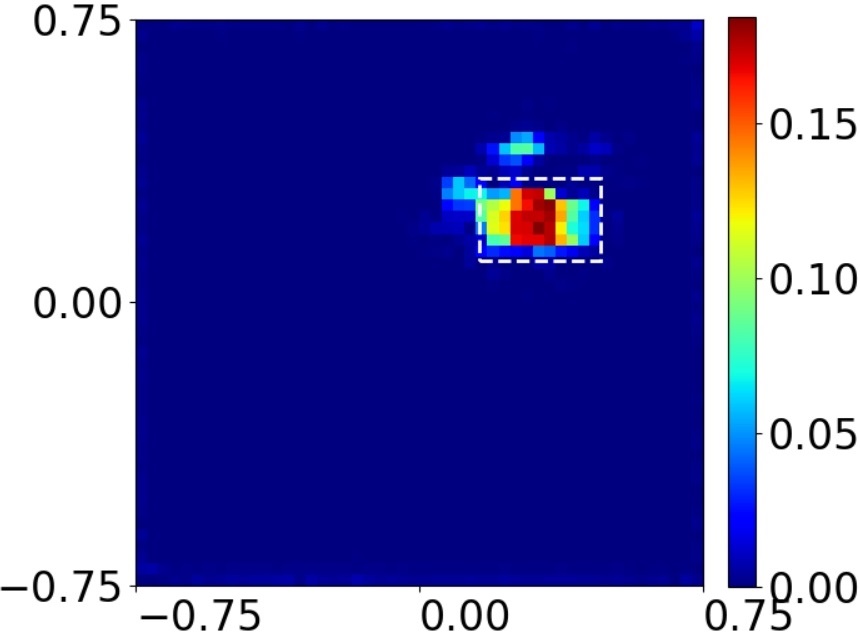}
			\subcaption{Stack of books with $\epsilon_r = 3.4 + 0.25j$}
		\end{subfigure}
		\begin{subfigure}{0.5\textwidth}
			\centering
			\includegraphics[scale=0.127]{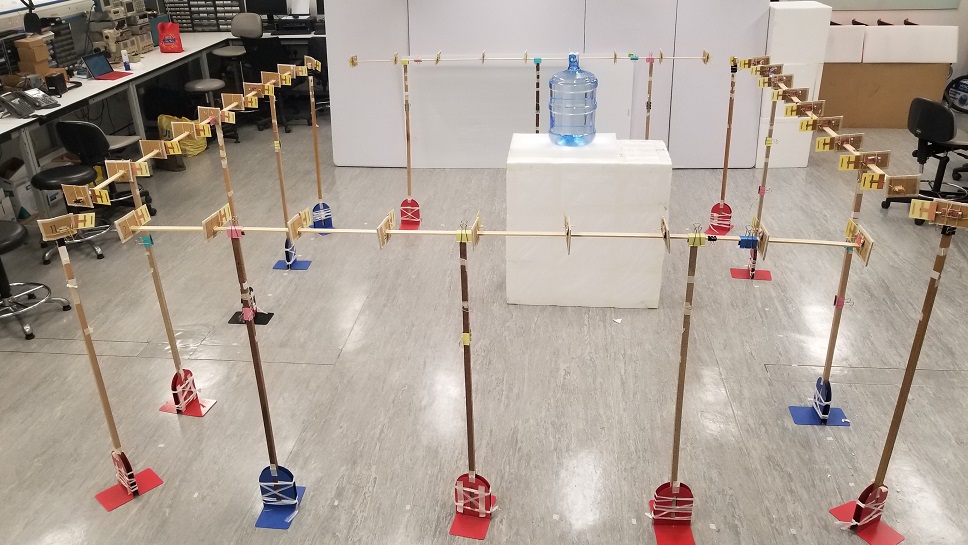}
			\includegraphics[scale=0.18]{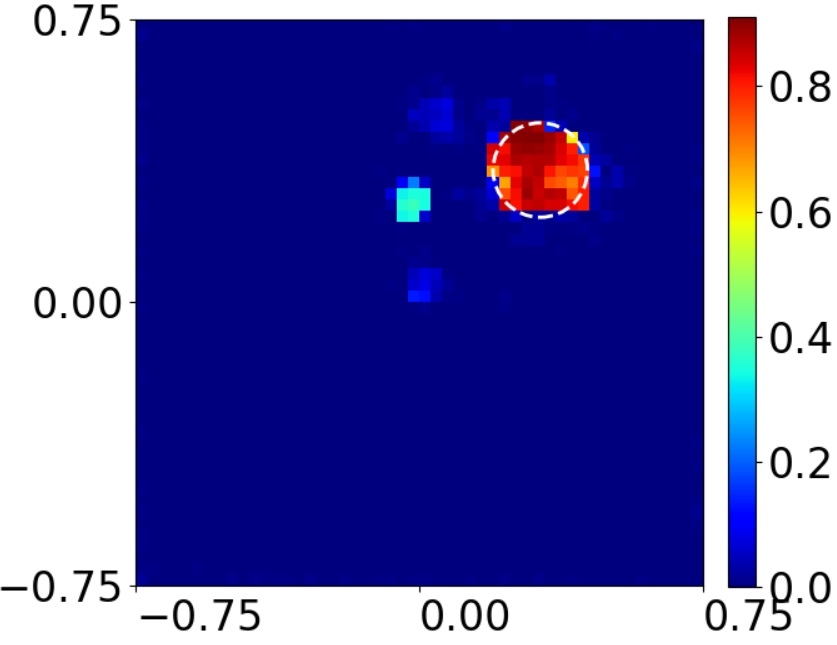}
			\subcaption{Water container with $\epsilon_r = 77 + 7j$}
		\end{subfigure}
		\begin{subfigure}{0.5\textwidth}
			\centering
			\includegraphics[scale=0.127]{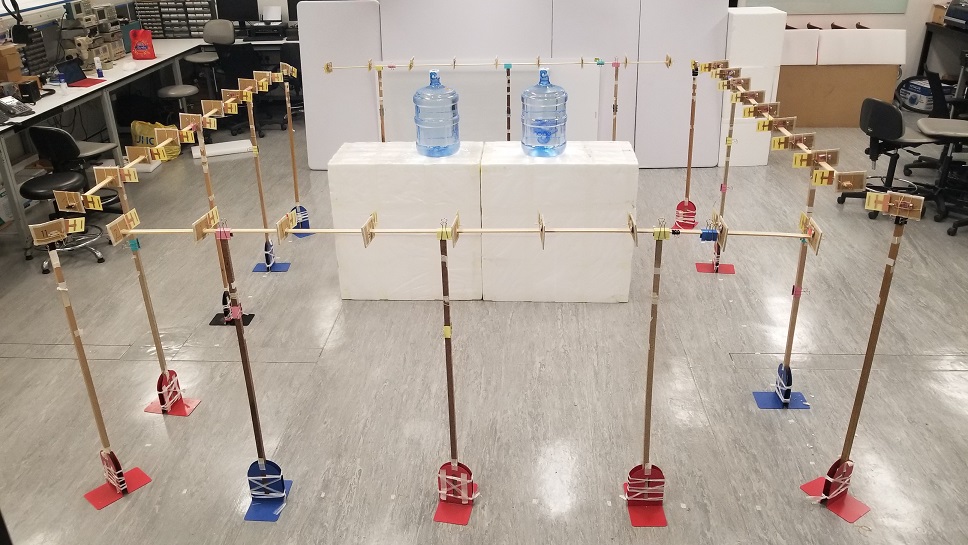}
			\includegraphics[scale=0.18]{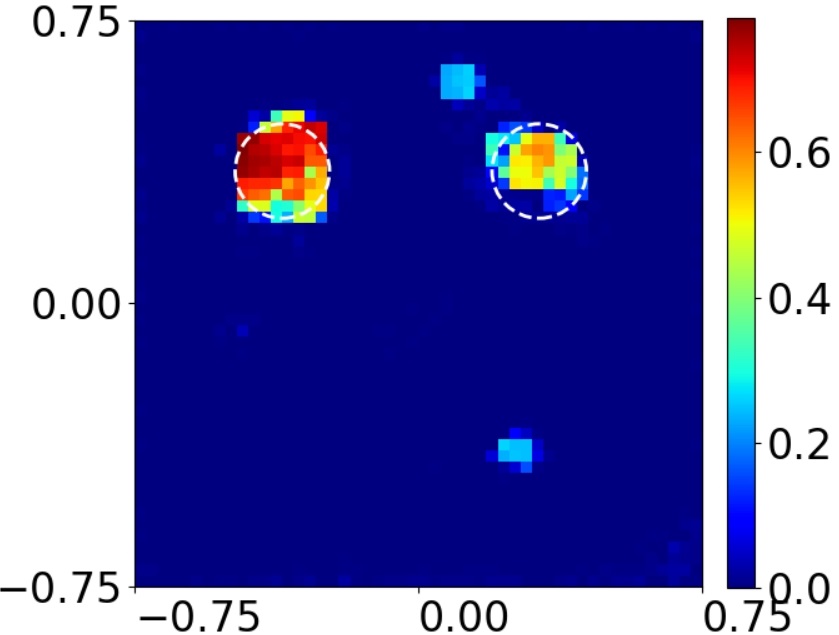}
			\subcaption{Two water containers with $\epsilon_r = 77 + 7j$ for both}
		\end{subfigure}
		\caption{Reconstruction results for various objects placed in the DoI wherethe  reconstructed $\delta(\bm{r}_n) \sqrt{\epsilon_R(\bm{r}_n)}$ values are indicated by color. First column shows the experiment ground truth and the second column shows the reconstruction using xPRA-TK-DPrior. The respective PSNR values of the reconstruction are (a) 36.42 dB, (b) 24.92 dB, and (c) 20.57 dB}
		\label{Results_exp_1}
	\end{figure}
	
	\begin{figure}[t]
		\centering
		\includegraphics[scale=0.16]{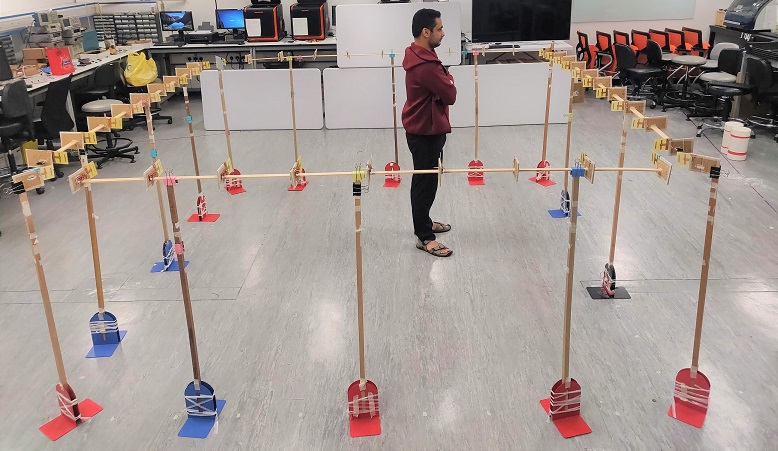}
		\includegraphics[scale=0.18]{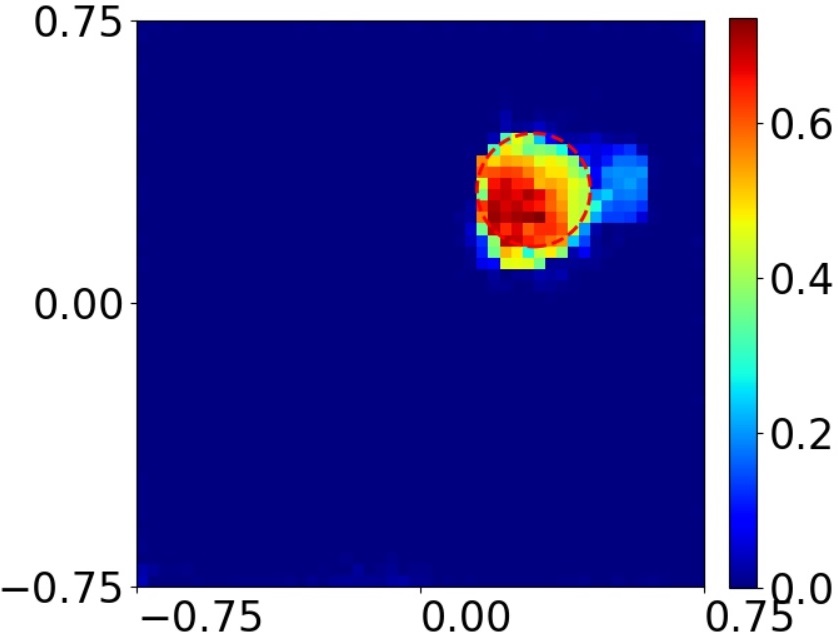}
		\caption{Reconstruction results for a person standing in the DoI where the reconstructed $\delta(\bm{r}_n) \sqrt{\epsilon_R(\bm{r}_n)}$ values are indicated by color. First figure shows the experiment ground truth and the second figure shows the reconstruction obtained using xPRA-TK-DPrior.}
		\label{Results_human}
	\end{figure}
	
	Fig. \ref{Results_exp_1} shows results for various such objects placed in the DoI. For all these results, the first column shows these objects placed on Styrofoam platforms in the DoI. We obtain the incident power $\overline{P}_i$ by only placing the Styrofoams in the DoI and then measure the total power $\overline{P}$ after placing the objects on the Styrofoam. We then obtain the measurement vector $\Delta \overline{P} = \overline{P} - \overline{P}_i$[dB], which contains the change in measurements due to the object being placed in the DoI, thereby cancelling the scattering effects due to the floor, ceiling and all the other clutter inside and outside the DoI. We then use these measurements to obtain reconstructions. Note that this straightforward background subtraction is only possible with the xPRA model due to its linear relationship between change in RSS values and contrast.
	
	Fig. \ref{Results_exp_1}(a) shows the reconstruction for a stack of books with $\epsilon_r = 3.4 + 0.25j$ (as measured by a high precision cavity resonator), so the value of $\delta \sqrt{\epsilon_R}$ is $0.135$. We can see that our framework is able to reconstruct the shape and $\delta \sqrt{\epsilon_R}$ value of this stack of books with a high degree of accuracy. Similarly, Fig. \ref{Results_exp_1}(b) and (c) show one and two containers of water being placed in the DoI respectively, for each of which $\epsilon_r = 77 + 7j$ leading to a $\delta \sqrt{\epsilon_R}$ value of $0.877$. Therefore, these are very strong scatterers for which most existing reconstruction techniques fail to work. However, for both these cases our framework is able to accurately localize the objects and reconstruct the $\delta(\bm{r}_n) \sqrt{\epsilon_R(\bm{r}_n)}$ values in the DoI. This can be seen through the PSNR values of the reconstructions mentioned in the figure caption.
	
	Fig. \ref{Results_human} shows the result for a person standing in the DoI. Various parts of the human body can have $\epsilon_R$ values between $50$ and $77$. Even for a human, our technique is able to accurately locate the person in the DoI and give accurate reconstructions of shape and $\delta(\bm{r}_n) \sqrt{\epsilon_R(\bm{r}_n)}$ values. Therefore, our framework can be efficiently used in indoor imaging applications.

	\section{Conclusion}
	
	An unrolled optimization framework with deep-learning based priors is proposed to solve highly non-linear and ill-posed phaseless data inverse scattering problems. The framework uses an approximate linear physics-based model along with data-driven regularization in the form of deep priors, which help extend the validity of the physics-based model to very strong scattering conditions. The optimum values of all tunable parameters are learned from data, thus removing the need to tune them manually. Results are demonstrated for both simulated and experimental data and show that the framework provides accurate reconstructions for objects exhibiting extremely strong scattering, while also generalizing to unseen shapes and permittivity values.
	
	\bibliographystyle{IEEEtran}
	\bibliography{refs}

\begin{thebibliography}{10}
\providecommand{\url}[1]{#1}
\csname url@samestyle\endcsname
\providecommand{\newblock}{\relax}
\providecommand{\bibinfo}[2]{#2}
\providecommand{\BIBentrySTDinterwordspacing}{\spaceskip=0pt\relax}
\providecommand{\BIBentryALTinterwordstretchfactor}{4}
\providecommand{\BIBentryALTinterwordspacing}{\spaceskip=\fontdimen2\font plus
\BIBentryALTinterwordstretchfactor\fontdimen3\font minus
  \fontdimen4\font\relax}
\providecommand{\BIBforeignlanguage}[2]{{%
\expandafter\ifx\csname l@#1\endcsname\relax
\typeout{** WARNING: IEEEtran.bst: No hyphenation pattern has been}%
\typeout{** loaded for the language `#1'. Using the pattern for}%
\typeout{** the default language instead.}%
\else
\language=\csname l@#1\endcsname
\fi
#2}}
\providecommand{\BIBdecl}{\relax}
\BIBdecl

\bibitem{chen2018computational}
X.~Chen, \emph{Computational methods for electromagnetic inverse
  scattering}.\hskip 1em plus 0.5em minus 0.4em\relax Wiley Online Library,
  2018.

\bibitem{xu2020deep}
K.~Xu, L.~Wu, X.~Ye, and X.~Chen, ``Deep learning-based inversion methods for
  solving inverse scattering problems with phaseless data,'' \emph{IEEE
  Transactions on Antennas and Propagation}, vol.~68, no.~11, pp. 7457--7470,
  2020.

\bibitem{chen2010subspace}
L.~Pan, Y.~Zhong, X.~Chen, and S.~P. Yeo, ``Subspace-based optimization method
  for inverse scattering problems utilizing phaseless data,'' \emph{IEEE
  Transactions on Geoscience and Remote Sensing}, vol.~49, no.~3, pp. 981--987,
  2010.

\bibitem{dubey2022xPRA}
A.~Dubey, S.~Deshmukh, L.~Pan, X.~Chen, and R.~Murch, ``A phaseless extended
  {R}ytov approximation for strongly scattering low-loss media and its
  application to indoor imaging,'' \emph{IEEE Transactions on Geoscience and
  Remote Sensing}, pp. 1--1, 2022.

\bibitem{jin2017deep}
K.~H. Jin, M.~T. McCann, E.~Froustey, and M.~Unser, ``Deep convolutional neural
  network for inverse problems in imaging,'' \emph{IEEE Transactions on Image
  Processing}, vol.~26, no.~9, pp. 4509--4522, 2017.

\bibitem{deshmukh2022physics}
S.~Deshmukh, A.~Dubey, D.~Ma, Q.~Chen, and R.~Murch, ``Physics assisted deep
  learning for indoor imaging using phaseless {Wi-Fi} measurements,''
  \emph{IEEE Transactions on Antennas and Propagation}, 2022.

\bibitem{sanghvi2019embedding}
Y.~Sanghvi, Y.~Kalepu, and U.~K. Khankhoje, ``Embedding deep learning in
  inverse scattering problems,'' \emph{IEEE Transactions on Computational
  Imaging}, vol.~6, pp. 46--56, 2019.

\bibitem{zhang2017learning}
K.~Zhang, W.~Zuo, S.~Gu, and L.~Zhang, ``Learning deep {CNN} denoiser prior for
  image restoration,'' in \emph{Proceedings of the IEEE conference on computer
  vision and pattern recognition}, 2017, pp. 3929--3938.

\bibitem{rick2017one}
J.~Rick~Chang, C.-L. Li, B.~Poczos, B.~Vijaya~Kumar, and A.~C.
  Sankaranarayanan, ``One network to solve them all - solving linear inverse
  problems using deep projection models,'' in \emph{Proceedings of the IEEE
  International Conference on Computer Vision}, 2017, pp. 5888--5897.

\bibitem{meinhardt2017learning}
T.~Meinhardt, M.~Moller, C.~Hazirbas, and D.~Cremers, ``Learning proximal
  operators: Using denoising networks for regularizing inverse imaging
  problems,'' in \emph{Proceedings of the IEEE International Conference on
  Computer Vision}, 2017, pp. 1781--1790.

\bibitem{diamond2017unrolled}
S.~Diamond, V.~Sitzmann, F.~Heide, and G.~Wetzstein, ``Unrolled optimization
  with deep priors,'' \emph{arXiv preprint arXiv:1705.08041}, 2017.

\bibitem{adler2017solving}
J.~Adler and O.~{\"O}ktem, ``Solving ill-posed inverse problems using iterative
  deep neural networks,'' \emph{Inverse Problems}, vol.~33, no.~12, p. 124007,
  2017.

\bibitem{schlemper2017deep}
J.~Schlemper, J.~Caballero, J.~V. Hajnal, A.~N. Price, and D.~Rueckert, ``A
  deep cascade of convolutional neural networks for dynamic {MR} image
  reconstruction,'' \emph{IEEE transactions on Medical Imaging}, vol.~37,
  no.~2, pp. 491--503, 2017.

\bibitem{Mittra1998}
A.~F. Peterson, S.~L. Ray, and R.~Mittra, \emph{Computational methods for
  electromagnetics}.\hskip 1em plus 0.5em minus 0.4em\relax IEEE press New
  York, 1998, vol. 351.

\bibitem{tibshirani2005sparsity}
R.~Tibshirani, M.~Saunders, S.~Rosset, J.~Zhu, and K.~Knight, ``Sparsity and
  smoothness via the fused lasso,'' \emph{Journal of the Royal Statistical
  Society: Series B (Statistical Methodology)}, vol.~67, no.~1, pp. 91--108,
  2005.

\bibitem{parikh2014proximal}
N.~Parikh and S.~Boyd, ``Proximal algorithms,'' \emph{Foundations and Trends in
  optimization}, vol.~1, no.~3, pp. 127--239, 2014.

\bibitem{boyd2011distributed}
S.~Boyd, N.~Parikh, E.~Chu, B.~Peleato, J.~Eckstein \emph{et~al.},
  ``Distributed optimization and statistical learning via the alternating
  direction method of multipliers,'' \emph{Foundations and
  Trends{\textregistered} in Machine learning}, vol.~3, no.~1, pp. 1--122,
  2011.

\bibitem{monga2021algorithm}
V.~Monga, Y.~Li, and Y.~C. Eldar, ``Algorithm unrolling: {I}nterpretable,
  efficient deep learning for signal and image processing,'' \emph{IEEE Signal
  Processing Magazine}, vol.~38, no.~2, pp. 18--44, 2021.

\bibitem{unetoriginal}
O.~Ronneberger, P.~Fischer, and T.~Brox, ``U-net: Convolutional networks for
  biomedical image segmentation,'' in \emph{International Conference on Medical
  image computing and computer-assisted intervention}.\hskip 1em plus 0.5em
  minus 0.4em\relax Springer, 2015, pp. 234--241.

\bibitem{deepimageprior}
D.~Ulyanov, A.~Vedaldi, and V.~Lempitsky, ``Deep image prior,'' in
  \emph{Proceedings of the IEEE conference on computer vision and pattern
  recognition}, 2018, pp. 9446--9454.

\bibitem{4562803}
Y.~Pinhasi, A.~Yahalom, and S.~Petnev, ``Propagation of ultra wide-band signals
  in lossy dispersive media,'' in \emph{2008 IEEE International Conference on
  Microwaves, Communications, Antennas and Electronic Systems}, 2008, pp.
  1--10.

\bibitem{ahmad2014partially}
F.~Ahmad, M.~G. Amin, and T.~Dogaru, ``Partially sparse imaging of stationary
  indoor scenes,'' \emph{EURASIP Journal on Advances in Signal Processing},
  vol. 2014, no.~1, pp. 1--15, 2014.

\bibitem{Productnote}
K.~C. Yaw, ``Measurement of dielectric material properties,'' \emph{Application
  Note. Rohde \& Schwarz}, pp. 1--35, 2012.

\bibitem{dubey2021enhanced}
A.~Dubey, P.~Sood, J.~Santos, D.~Ma, C.-Y. Chiu, and R.~Murch, ``An enhanced
  approach to imaging the indoor environment using {WiFi} {RSSI}
  measurements,'' \emph{IEEE Transactions on Vehicular Technology}, vol.~70,
  no.~9, pp. 8415--8430, 2021.

\end{thebibliography}
	
\end{document}